\documentclass[a4paper,11pt]{article}

\pdfoutput=1 

\usepackage{jinstpub} 

\usepackage{siunitx}

\title{Simulation study for an in-situ calibration system for the measurement of the snow accumulation and the index-of-refraction profile for radio neutrino detectors}
\author[a]{J. Beise}
\author[a]{C. Glaser}

\affiliation[a]{Uppsala University Department of Physics and Astronomy, SE-75237, Sweden}

\emailAdd{jakob.beise@physics.uu.se, christian.glaser@physics.uu.se}

\abstract{Sensitivity to ultra-high-energy neutrinos ($E>$\SI{e17}{eV}) can be obtained cost-efficiently by exploiting the Askaryan effect in ice, where a particle cascade induced by the neutrino interaction produces coherent radio emission that can be picked up by antennas. As the near-surface ice properties change rapidly within the upper $\mathcal{O}$(\SI{100}{m}), a good understanding of the ice properties is required to reconstruct the neutrino properties. In particular, continuous monitoring of the snow accumulation (which changes the depth of the antennas) and the index-of-refraction $n(z)$ profile are crucial for an accurate determination of the neutrino's direction and energy. We present an in-situ calibration system that extends the radio detector station with two radio emitters to continuously monitor the firn properties within the upper \SI{40}{m} by measuring the time differences between direct and reflected (off the surface) signals (D'n'R). We determine the optimal positions of two transmitters at all three sites of current and future in-ice radio detectors: Greenland, Moore's Bay, and the South Pole. For the South Pole we find that the snow accumulation $\Delta h$ can be measured with a resolution of \SI{3}{mm} and the parameters of an exponential $n(z)$ profile $\alpha$ and $z_0$ with $0.04\%$ and $0.14\%$ precision respectively, which constitutes an improvement of more than a factor of 10 as compared to the inference of the $n(z)$ profile from density measurements. 
Additionally, as this technique is based on the measurement of the signal propagation times we are not bound to the conversion of density to index-of-refraction. 
We quantify the impact of these ice uncertainties on the reconstruction of the neutrino vertex, direction, and energy and find that the calibration device measures the ice properties to sufficient precision to have negligible influence. 
}


\arxivnumber{2205.00726} 

\begin{document}
\maketitle
\flushbottom

\section{Introduction}
\label{sec:intro}

Ultra-high-energy (UHE) neutrino physics ($E>$ \SI{e17}{eV}) constitutes a unique tool to examine the most violent processes of our Universe \cite{Ackermann2019, Ackermann:2019cxh}. As neutrinos interact only via the weak force they can traverse through dense astrophysical environments almost unaffected allowing them to probe physical regions unreachable with gamma-rays, which for energies above 1~PeV are suppressed due to interactions with low-energetic radiation fields \cite{Franceschini2008}. UHE neutrinos are produced by interactions of cosmic rays (CR) with surrounding matter in proximity to their origin (astrophysical neutrinos) or radiation fields  during their propagation to Earth (cosmogenic neutrinos) \cite{Beresinsky1969}, but unlike CRs, they are not deflected by magnetic fields during their journey \cite{Aharonian2011}. Detecting UHE neutrinos thereby poses a smoking gun signature for the production sites and gives valuable insights into the production mechanisms of the most energetic particles in our Universe.

Because of the comparatively low neutrino interaction cross section, large detection volumes are required for a reasonable event rate. The IceCube detector \cite{Aartsen2017} at the Amundsen-Scott South Pole Station in Antarctica, composed of 5160 optical sensors, so-called Digital Optical Modules, deployed in ice at a depth between \SI{1450}{m} to \SI{2450}{m} is the currently largest neutrino observatory. The \SI{1}{km^3} of instrumented ice acts both as the interaction and detection medium. 
Recent IceCube data shows that the astrophysical muon neutrino flux follows a single power-law energy spectrum with $\Phi \propto E^{-\gamma}$ with the spectral index $\gamma$ between $2.37$ and $2.87$ \cite{MuNuFlux2021, HESENu2021} eventually falling below sensitivity for energies above $10^{16}$~eV. For the prospective IceCube-Gen2 high-energy detector extension, a significant sensitivity improvement is envisaged for energies up to \SI{e18}{eV} \cite{Aartsen2021}, which can only be achieved by covering an effective volume of $\mathcal{O}(100)$ larger than the current detector. The most cost-efficient technique to cover such effective volumes is provided by the in-ice radio detection technique \cite{BarwickGlaser2022}. This has to do with the comparatively large attenuation length of radio signals in cold ice of $\mathcal{O}$(\SI{1}{km}) \cite{Barwick2005} making it possible to effectively cover large volumes with few stations.

The radio emission is generated via the Askaryan effect \cite{Askaryan1962, Alvarez1998, Alvarez2000}, as a result of a moving, time-varying charge excess during the propagation of the charged secondaries of a shower. The coherent radio flashes are strongest close to the Cherenkov angle in a frequency range between \SI{50}{MHz} to \SI{1}{GHz}.

In-ice radio neutrino detectors are composed of an array of individual and autonomous radio stations spread over large distances to optimize the total effective area. Today, there are two existing pilot projects: ARIANNA at the Ross Ice Shelf and the South Pole \cite{Anker2019} and ARA at the South Pole \cite{Allison2016}. RNO-G \cite{Aguilar2021}, currently under construction at Summit Station, Greenland, and the proposed ARIANNA-200 detector \cite{Anker2020}, are detectors of similar sensitivity but different sky coverage. Their sensitivity is large enough that the first measurement of a UHE neutrino seems possible in the next years. IceCube-Gen2 also features a radio array \cite{Hallmann2021} which will improve the sensitivity to UHE neutrino by another order of magnitude. 

Generally, two different radio station designs are considered for radio arrays varying in the depth of the deployed antennas. This study focuses on a shallow design, which is the proposed layout for the ARIANNA-200 and part of the layout of RNO-G and the radio detector of IceCube-Gen2. The shallow design comprises upwards- and downwards-facing log-periodic dipole antennas (LPDAs) in \SI{3}{m} depth and a vertically polarised cylindrical omni-directional antenna (VPol) buried at approx. \SI{15}{m} depth. Typical choices are bicone or fat dipole antennas. The deep Vpol antenna will for almost all neutrino events record two signals: a direct signal propagating on a curved trajectory and a slightly time-delayed second pulse reflected off the ice-air interface. This so-called D'n'R (direct and reflected) signature, in particular the time difference between the two pulses (D'n'R time difference), can be used as a proxy for the neutrino vertex distance and to discriminate in-ice radio signals from background \cite{DnR2019}. The depth of the Vpol antenna is a compromise between resolution (the deeper the better), fraction of events that show a D'n'R signature (the shallower the better), and deployment effort (the shallower the better). In this work, we fix the depth to \SI{15}{m} but as discussed in Ref.~\cite{DnR2019} also shallower depths up to \SI{10}{m} still show two clearly separated pulses and would yield an energy resolution below the natural limit from inelasticity fluctuations of a factor of two. 

The density of the near-surface ice (also referred to as firn) changes rapidly from loose snow at the surface to compact ice within $\mathcal{O}$(\SI{100}{m}). This affects the index-of-refraction $n$ and thereby the propagation of the radio signals. With a depth-dependent density gradient leading to approximately $n=1.35$ at the surface to $n=1.78$ for deep ice, rays that would otherwise travel upwards on a straight line are bent downwards to a curved trajectory. In addition, the accumulation of fresh snow effectively ``buries'' the antennas deeper which delays the arrival of the reflected pulse. In order to correctly reconstruct the neutrino vertex and energy from the D'n'R signature, a precise knowledge of the firn properties, including the index-of-refraction profile and the snow accumulation, is essential. The parameters have to be determined with high enough accuracy and monitored continuously (in case of the snow accumulation) in order to take temporal variations into account. 

\begin{figure}[tbp]
        \centering
        \includegraphics[width = \textwidth]{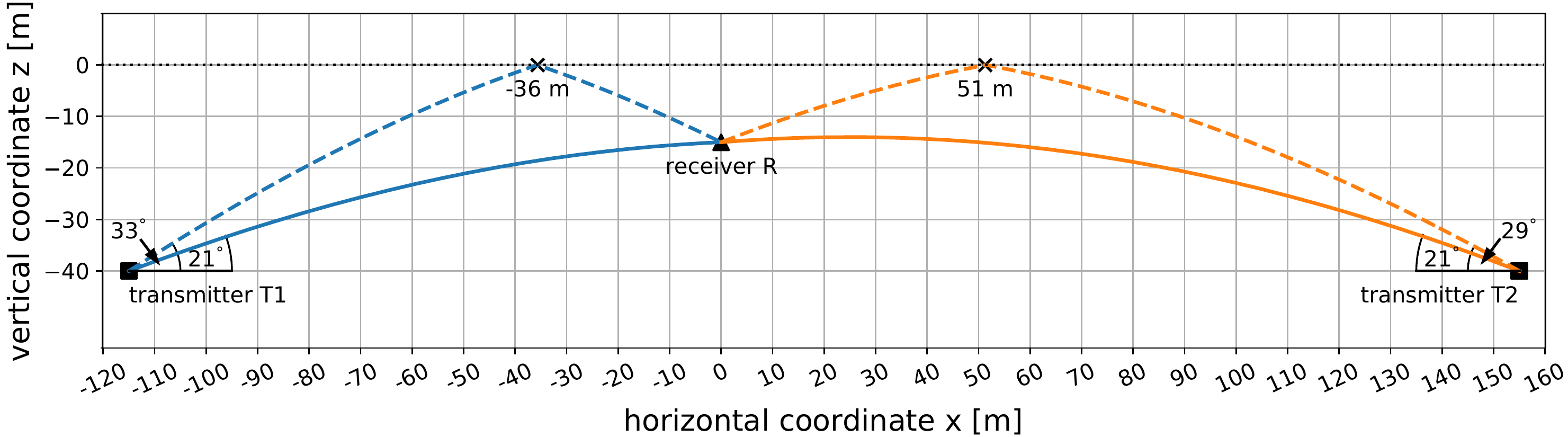}
        \caption{Sketch of a calibration system consisting of two transmitters (T1 and T2, black square) and one receiver (R, black triangle) antenna (for a detailed discussion of the choice of antenna positions we refer to Sec.~\ref{sec:cal_sim}). A \SI{15}{m} deep Vpol antenna picks up the D’n’R signal composed of the early direct signal (solid line) and a slightly time-delayed, second pulse reflected off the ice-air interface (dashed line). The time delay between both signals relates to the snow accumulation. We show the optimal transmitter positions for a detector at the South Pole (see Sec.~\ref{sec:cal_result}) with T1~=~[-115,~-40]~m and T2~=~[155,~-40]~m.}
        \label{fig:southpole}
    \end{figure}

So far, the index-of-refraction $n(z)$ profile has been derived from density measurements of ice cores (see Ref.~\cite{Barwick2018} for an overview of available measurements) assuming an empiric conversion factor between ice density and index-of-refraction. In this work, we present a calibration system that measures the $n(z)$ profile directly through propagation times of radio waves in ice, i.e., the relevant property for radio neutrino detectors. Furthermore, the system can be set up for continuous measurements (e.g. one measurement every 12 hours) to track the snow accumulation and any temporal changes of the $n(z)$ profile. The caveat of the calibration system is that we do not measure the numerical value of $n$ per depth $z$ but fit a functional form of $n(z)$ to the propagation time measurements.  
A prototype calibration system for the reconstruction of the snow accumulation consisting of a single transmitter-receiver pair  was already deployed at the ARIANNA site on the Ross Ice Shelf, Antarctica in 2018 \cite{DnR2019} and has been operating ever since. The study showed the experimental feasibility of such a calibration system, in particular it was shown that the propagation times from transmitter to receiver could be measured with high precision and that all potential background could be rejected or excluded (see \cite{DnR2019} for details). Building upon this, we investigate an in-situ calibration system comprising two transmitter antennas, capable of simultaneously determining both the snow accumulation and the index-of-refraction profile (Fig.~\ref{fig:southpole}). Two transmitters are the minimal number to measure the $n(z)$ profile and we restrict this work to this choice to minimize deployment efforts but additional transmitters are foreseeable in the future to increase precision and add redundancy to the measurement. We note that due to reciprocity, the calibration device can also be built with two receivers and one transmitter but such a setup will likely be more expensive and will have larger uncertainties because it requires the measurement of time delays between signals received in two antennas compared to time delays within a single waveform. 

The focus of this paper relies on the application for IceCube-Gen2 Radio at the South Pole and is structured as follows: In Sec.~\ref{sec:cal} we optimize the calibration setup in a simulation study and study systematic uncertainties. In Sec.~\ref{sec:nuprop} we study the implications of the in-situ calibration capabilities on the reconstruction of the neutrino vertex distance, the neutrino direction, and the neutrino energy for variations in the parameters of the ice model (Sec.~\ref{sec:nuprop_iceprop}) and the snow height (Sec.~\ref{sec:nuprop_snowheight}). In Sec.~\ref{sec:conclusion} we summarize our results and put them in a broader context.

\section{In-situ Calibration for the Measurement of Ice Properties}
\label{sec:cal}

In this section, we first present the modelling of the index-of-refraction profile (Sec.~\ref{sec:nz}), details of the simulation study (Sec.~\ref{sec:cal_sim}), the achievable resolution (Sec.~\ref{sec:cal_result}), and discuss sources of systematic uncertainties (Sec.~\ref{sec:cal_syst}). We note that we have repeated the simulation also for a setup on Greenland (RNO-G, appendix~\ref{sec:greenland}) and for the Ross Ice Shelf (ARIANNA-200, appendix~\ref{sec:mooresbay}) which differ from one another in the initial parameters of the index-of-refraction profile (see Tab.~\ref{tab:model}). This section presents the simulation for the South Pole site.

\subsection{The index-of-refraction profile}
\label{sec:nz}

The firn properties, relevant for the reconstruction of the neutrino energy and direction, are the snow accumulation and the index-of-refraction profile. The former is defined as the relative increase $\Delta h$ of the snow height relative to a reference point, whereas the latter, the depth-dependent index-of-refraction profile $n(z)$, can be modeled by an exponential function \cite{Barwick2018} of the form

\begin{equation}
    \centering
    \begin{split}
    n(z) & = 1.78 - \Delta n \cdot e^{-(z+\Delta h)/z_0},
    \end{split}
    \label{eq:icemodel1}
\end{equation}

\noindent with $z$ being the depth, 1.78 the refractive index of deep ice \cite{Warren1984}, and $z_0$ the characteristic length. The parameter $\Delta n$ specifies the index-of-refraction at the surface of the firn $n_{z=0} = 1.78 - \Delta n$. 

We note that there are indications that a single exponential function cannot describe the index-of-refraction profile over the full depth range from the surface to deep in the ice \cite{Allison2022} and that a double exponential or an even more complex model might be required to improve the modeling precision over the full depth range \cite{Deaconu2018,PhDThesisLatif}. Theoretically, the transition points are expected at the critical densities of \SI{550}{kg/m^3} and \SI{830}{kg/m^3} \cite{tc-16-2683-2022} which are at approx.~\SI{20}{m} and \SI{120}{m} at the South Pole. Data from the calibration system we propose in this article can also be used to check for deviations from a single exponential model. 
In addition, the transition with depth is not perfectly smooth but fluctuations around the smooth profile exist which lead to second order propagation effects such as the existence of potentially detectable (though generally small) signals coming from regions from where no classical propagation paths exist \cite{Barwick2018}, diffraction and interference of the radio waves, and the presence of caustics, where the small electric field may be significantly amplified in some geometries \cite{Deaconu2018,RadarEchoTelescope:2020nhe}.
However, an exponential $n(z)$ profile provides a good description of the index-of-refraction data points derived from density measurements with residuals below 1$\%$ \cite{Barwick2018}. Furthermore, detailed measurements of the signal arrival directions at the surface of radio pulses emitted deep in the ice showed that a single exponential $n(z)$ profile is sufficient to correct the bending of the signal trajectories to sub-degree precision \cite{ARIANNA:2020zrg}.

We further note that birefringence was found to have an important effect at the South Pole \cite{Jordan, Amy, Heyer2022}, i.e., that the index-of-refraction depends on the polarization and propagation direction of the radio pulse. The birefringence asymmetry is at the per-mill level at the South Pole and even smaller at the other sites considered for in-ice radio detection of neutrinos. Therefore, birefringence can safely be ignored in this study, as we deal with fairly short propagation lengths close to the surface where the birefringence asymmetry is smallest.

The goal of the in-situ calibration system is to measure the parameters $\Delta h$, $\Delta n$ and $z_0$. However, it can be shown that there are only two effective parameters $\alpha$ and $z_0$ by reformulating Eq.~\eqref{eq:icemodel1} to:

\begin{equation}
    \centering
    \begin{split}
    n(z) & = 1.78 - \alpha \cdot e^{-z/z_0},\\
    \end{split}
    \label{eq:icemodel2}
\end{equation}
with the new effective parameter
\begin{equation}
    \alpha  = \exp{(\ln{(\Delta n)} -\Delta h / z_0 )}. 
    \label{eq:alpha}
\end{equation}

This means that snow accumulation, i.e., an increase in $\Delta h$, has the same effect as changing the parameter $\Delta n$. The two parameters that can be reconstructed simultaneously are $\alpha$ and $z_0$. There are several options to disentangle snow accumulation from a change in the $\Delta n$ parameter. 
\begin{itemize}
    \item For many analyses, only the index-of-refraction profile $n(z)$ is needed and it is not important if a change originated from a change in $\Delta n$ or $\Delta h$. Therefore, a measurement of the effective parameter $\alpha$ and $z_0$ is sufficient.
    \item The parameter $\Delta n$ is directly related to the index-of-refraction at the surface ($n_{z=0} = 1.78 - \Delta n$) which can be determined by measuring the snow density at the surface and exploiting the relation between density and index-of-refraction. Furthermore, it is not expected that this parameter changes significantly over time. Then, for a fixed $\Delta n$, the parameters $\Delta h$ and $z_0$ can be measured simultaneously. 
    \item The snow accumulation can be measured by other means, e.g., a simple meter stick that is read off by eye. In particular, after the initial deployment of the calibration system, the snow level is known and the parameters $\Delta n$ and $z_0$ can be measured. After that, it can be assumed that $\Delta n$ stays constant, and the snow accumulation can be tracked over time. Often, the site can be visited once a year and the snow accumulation can be read off from the meter stick again. This can be used to double check that the snow accumulation measurement from the calibration system was correct, and with known snow accumulation, the parameters $\Delta n$ and $z_0$ can be remeasured to probe any change in time. 
\end{itemize}

We find the last option most promising, but only future experimental tests will give a definitive answer. In the following, we will develop a calibration system to determine the parameters $\alpha$ and $z_0$ as precise as possible, according to Eq.~\ref{eq:icemodel2}.

\begin{table}[t]
\centering
\begin{tabular}{ccc}
\hline\hline
Site & $\Delta n^\mathrm{true}$ & $z_0^\mathrm{true}$ [m]\\
\hline
South Pole \cite{Barwick2018} & 0.423 & 77.0\\
Greenland \cite{Deaconu2018} & 0.51 & 37.25\\
Moore's Bay \cite{Barwick2018}  & 0.46 & 34.5\\
\hline\hline
\end{tabular}
\caption{\label{tab:model}Initial (true) values for the parameters of the change of refractive index $\Delta n$ and the characteristic length $z_0$ at the three geographical sites Moore's Bay, South Pole and Greenland.}
\end{table}

\subsection{Simulation Setup}
\label{sec:cal_sim}

For simplicity, the further study is reduced to two spacial dimensions, because we can always rotate around the vertical axis such that the signal propagates along the x-direction. We discuss possible tilts in the snow surface and surface roughness that would break the azimuthal symmetry later in Sec.~\ref{sec:cal_syst_fresnel}. The calibration system constitutes two transmitter and one receiver antenna. In operation, the transmitters would be periodically activated by a pulse generator emitting radio signals that would be picked up by the receiver (R) (see figure \ref{fig:southpole}). Having two transmitters deployed results in four trajectories, a direct and reflected path for each transmitter. To reconstruct two parameters requires at least two independent observables - here we use the three relative time differences between the propagation times. To calculate the propagation paths we use the ray-tracing technique (see e.g. \cite{NuRadioMC2019}). In order to find the propagation paths for a given transmitter configuration the initial set of parameters was defined to be $\alpha = \Delta n^\mathrm{true}$ = 0.423 and $z_0^\mathrm{true}$ = \SI{77.0}{m}. The objective of the simulation is to fit these values with highest precision utilizing solely the detector geometry and the three observables ($\Delta t_1,\ \Delta t_2,\ \Delta t_3$), which are defined by the propagation times of the direct $t_\mathrm{dir}$ and reflected $t_\mathrm{ref}$ pulses for transmitter 1 (T1) $t^\mathrm{T1}$ and transmitter 2 (T2) $t^\mathrm{T2}$ as follows:

\begin{align}
    \centering
    \nonumber \Delta t_1 & = t_\mathrm{ref}^\mathrm{T1} - t_\mathrm{dir}^\mathrm{T1} \ ,\\
    \nonumber \Delta t_2 & = t_\mathrm{dir}^\mathrm{T2} - t_\mathrm{dir}^\mathrm{T1} \ ,\\ 
    \Delta t_3 & = t_\mathrm{ref}^\mathrm{T2} - t_\mathrm{dir}^\mathrm{T1} \ .  
    \label{eq:observables}
\end{align}

\begin{figure}[tbp]
    \centering
    \includegraphics[width = \textwidth]{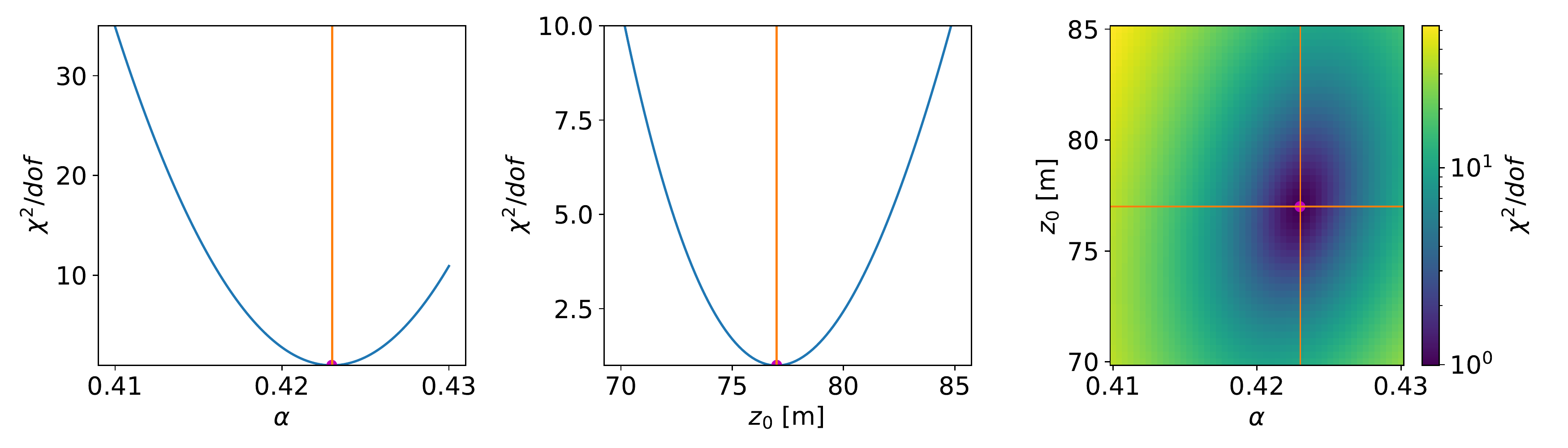}
    \caption{$\chi^2$-profile for $\alpha$ (left), $z_0$ (middle) and the two-dimensional $\chi^2$-contour $\alpha-z_0$ (right) at the position T1~=~[-115,~-40]~m and T2~=~[-155,~-40]m, which we later find to be the optimal configuration. The magenta-colored dot represents the optimised values, while the orange line shows the true value.}
    \label{fig:chi2fit}
\end{figure}

Imitating a real detection that underlies statistical fluctuations, a sample of 10,000 propagation times is randomly drawn from a Gaussian distribution centered around the true propagation time with a width of \SI{0.2}{ns}. The prototype study with a single transmitter showed that \SI{0.2}{ns} is achievable under realistic experimental conditions \cite{DnR2019}. The $3 \times 10,000$ observables are fitted at once to obtain $\alpha$ and $z_0$ using \nolinkurl{iminuit} \cite{Iminuit} - a \nolinkurl{Python} interface for the \nolinkurl{C++} implemented \nolinkurl{Minuit2} minimizer. Figure~\ref{fig:chi2fit} depicts the $\chi^2$-profiles in $\alpha$ and $z_0$, as well as the two-dimensional contour, for the two transmitters being at position T1~=~[-115,~-40]~m and T2~=~[-155,~-40]~m. The optimised parameters are marked by the magenta-colored dot, while the orange line represents the true parameter value. The $\chi^2$-contour $\alpha$-$z_0$ is an indicator for the correlation between the parameters, which in this case is low. This will be further discussed in the next section.

In order to evaluate how reliably a certain antenna configuration reconstructs the ice properties, an uncertainty estimate on the fitted parameters is required. Therefore, for each configuration, $\alpha$ and $z_0$ are reconstructed for $N_{rep}= 2,000$ random realisations of the $3 \times 10,000$ observables. The number of repetitions $N_{rep}$ was chosen to reduce the statistical uncertainties to an acceptable level. The estimated correlation between $\alpha$ and $z_0$ has the largest uncertainty. Using a bootstrap technique we determined that the uncertainty in correlation dropped from 6.9$\%$ to 2.2$\%$ to 1.4$\%$ with an increase in repetitions from 200 to 2,000 to 5,000. We decided that the gain by going from 2,000 to 5,000 realisations does not justify the increase in computing time and used $N_{rep} = 2000$ in the following. 

\begin{figure}[tbp]
    \centering
    \includegraphics[width = \textwidth]{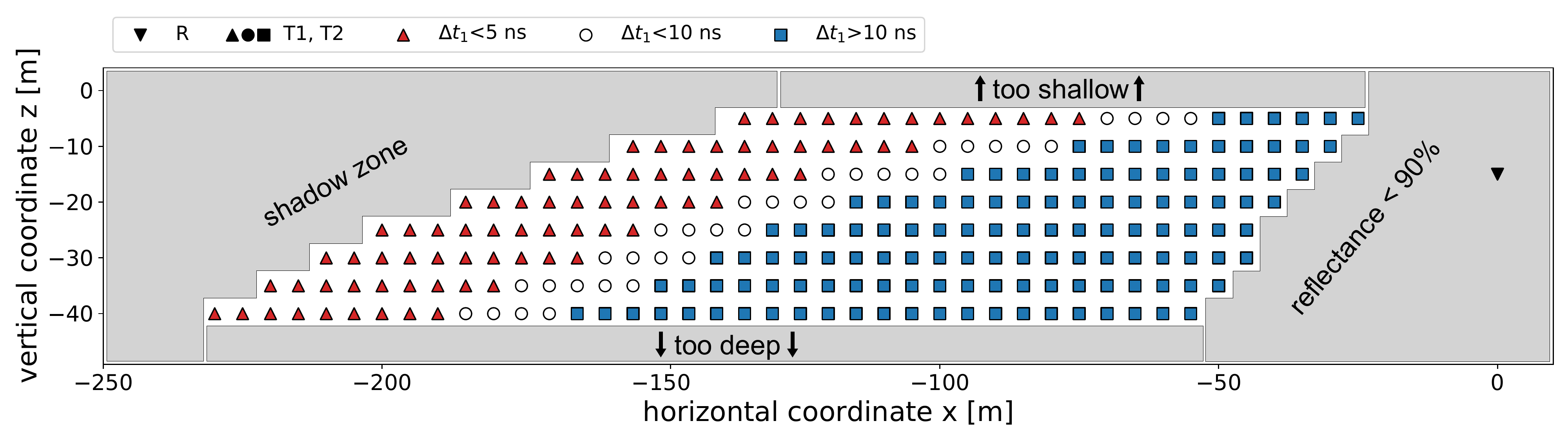}
    \caption{We only consider positions in the lower left quadrant of Fig.~\ref{fig:southpole} due to the rotational symmetry of the setup. The grid shows all possible positions of transmitter T1 and T2 as colored squares for the South Pole location for depths of -40~m. The position of the receiver R is indicated by the downwards-facing triangle on the right side at $x$=\SI{0}{m} and $z$=\SI{-15}{m}. The allowed transmitter locations are restrained by the shadow zone to the upper left and the 90$\%$ reflectance condition to the lower right. The antennas may not be buried too shallow (>\SI{-5}{m}) to ensure that the D'n'R signature is pronounced and not too deep (<\SI{-40}{m}) because of deployment constraints. For reconstruction purposes, the D'n'R pulses should be separated by more than \SI{10}{ns}. The colour code displays the D'n'R time difference with $\Delta t_1 <$ \SI{5}{ns} (red triangles), $\Delta t_1 <$ \SI{10}{ns} (white circles) and $\Delta t_1 >$ \SI{10}{ns} (blue squares).}
    \label{fig:grid_sp}
\end{figure}

Having the position [$x$,~$z$] of all antennas (R, T1, T2) as open parameters allows for many different configurations (Fig.~\ref{fig:grid_sp}). However, since the receiver antenna position is constrained by the neutrino energy reconstruction \cite{DnR2019}, a variation of this position is not considered in the study and thus fixed to R~=~[0,~-15]~m. For the two transmitters T1 and T2 a systematic way of testing all possible combinations is pursued by doing a grid scan over the allowed parameter space. Because of the symmetry of the setup only half of the transmitter space has to be studied, as a transmitter located at [$x$,~$z$] yields the same as one located at [-$x$,~$z$]. In this study we always consider the transmitters to be placed in the lower left quadrant with the receiver antenna in the upper right by the center of the coordinate frame (c.f. Fig.~\ref{fig:southpole}). Additionally, the transmitter space is restrained by the shadow zone (the region from which no two solutions reach the receiver) and by imposing a high reflectivity of 90$\%$ or higher for the reflected signal at the ice-air interface. The latter is imposed for the signal to have a strong enough amplitude for reconstruction, while the former ensures that there always are two ray-tracing solutions (cf. Fig.~\ref{fig:grid_sp}). 

Furthermore, the antennas should be deployed several wavelength underneath the ice (>\SI{-5}{m}). On the other hand, the depth is restricted to (<\SI{-40}{m}) due to deployment constraints and the technical limitations of the cylindrical melter that is foreseen to be used to melt the holes \cite{Heinen2017,Meltingprobe}. A depth of \SI{40}{m} is already field-proven and can be melted within a working day \cite{DnR2019}. Anyhow, for completion we also simulated deeper positions up to a depth of \SI{200}{m} by extending the above grid with a sparser spacing of \SI{20}{m} in $x$ and testing depths of [\SI{-50}{m}, \SI{-80}{m}, \SI{-100}{m}, \SI{-150}{m}, \SI{-200}{m}]. We also investigate restricting the maximum deployment depth to just \SI{20}{m} which would speed up the deployment time. We report these findings at the end of this section. 

The colour code in Fig.~\ref{fig:grid_sp} displays the D'n'R time difference ($\Delta t_1$ in Eq.~\ref{eq:observables}) measured at the respective transmitter position, categorized into those smaller than \SI{5}{ns} (red triangles), smaller than \SI{10}{ns} (white circles) and larger than \SI{10}{ns} (blue squares). We restrain ourselves to the positions exceeding a D'n'R time difference of \SI{10}{ns} as any realistic extraction of the D'n'R time difference will need to take the FWHM pulse width of approx.~\SI{2}{ns} into account which can cause interference effects between the pulses should they arrive in quick succession. Mixing of pulses originating from different transmitters can be separated by including a fixed cable delay so that we do not add this as an additional constraint here. 

For the South Pole site the remaining, truncated grid to depths of \SI{40}{m} has 127 positions in the x-z plane of 5 m by 5 m spacing ranging from $x \in [-25,~-165]$~m and $z \in [-5,~-40]$~m. It is worth noting that out of the $127 \times 126$ possible combinations of T1 and T2 there are only $(127 \times 126)/2$ unique combinations, reducing the computational effort by a factor of 2. This is due to the interchangeability of the two identical transmitters, that is (T1, T2) = (T2, T1).

The simulation produces 2,000 realisations for the entire grid of 127 positions. To evaluate which configuration is best to reconstruct the $n(z)$ profile we use the quadratic difference between true and reconstructed index-of-refraction profile as metric. As the deviation depends on the depth, we calculate the average deviation in the depth range of interest between \SI{-40}{m} and the surface (\SI{0}{m}) in steps of \SI{2}{m}. The metric is defined as

\begin{equation}
    \label{eq:metric}
    \centering
    \mathcal{M} = \sqrt{\frac{1}{N_{rep}N_{z}}\sum_{i=1}^{N_{rep}} \sum_{j=1}^{N_{z}} \left( n(z_j|\alpha^{true}, z_0^{true}) - n(z_j|\alpha^i, z_0^i)\right)^2 },
\end{equation}

where $n(z_j|\alpha^{true}, z_0^{true})$ is the refractive index $n$ at depth $z_j$ for the true $\alpha^{true}$ and $z_0^{true}$ and $n(z_j|\alpha^i, z_0^i)$ the refractive index $n$ at depth $z_j$ for realisation $i$ of $\alpha^i$ and $z_0^i$.  We also average over the 2000 repetitions to reduce the statistical uncertainties. 
$N_{rep}$ is the number of repetitions and $N_z$ is the number steps in depth, here 20. The metric $\mathcal{M}$ states how well the reconstructed $n(z)$ profile resembles the true profile in the critical first \SI{40}{m} of the firn. In addition to a low metric, we also aim to minimize the correlation $\rho_{\alpha, z_0}$ between $\alpha$ and $z_0$.

\begin{figure}[tbp]
    \centering
    \includegraphics[width=\textwidth]{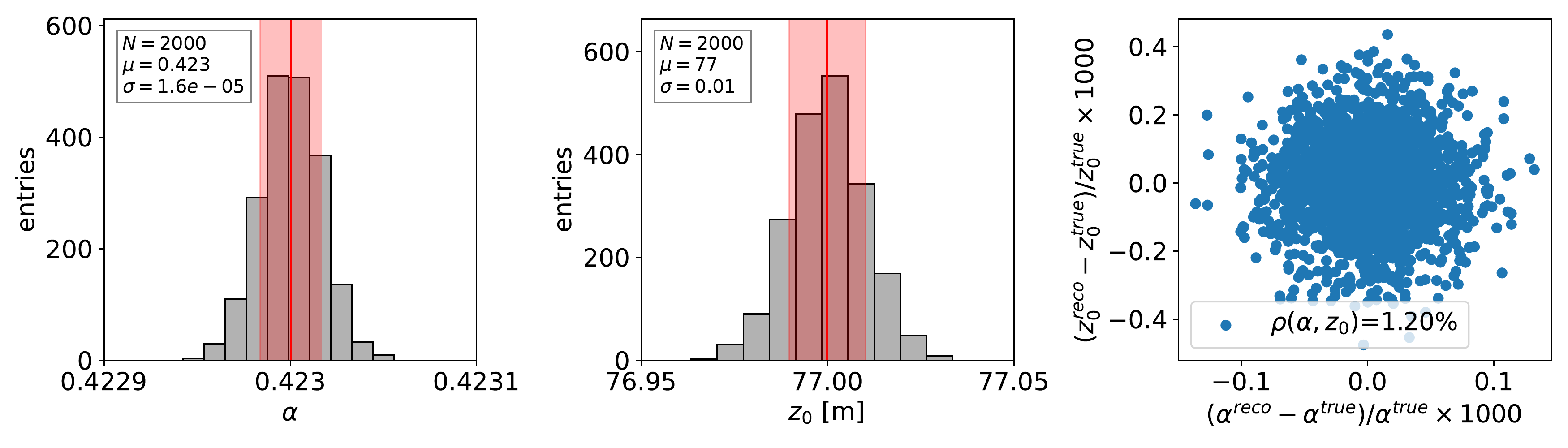}
    \caption{Parameter distribution in $\alpha$ and $z_0$ for 2,000 repetitions at position T1~=~[-115,~-40]~m and T2~=~[-155,~-40]~m. The red band indicates the 1$\sigma$ uncertainty interval.}
    \label{fig:iceprop_distribution}
\end{figure}

\begin{figure}[tbp]
    \centering
    \includegraphics[width=0.33\textwidth]{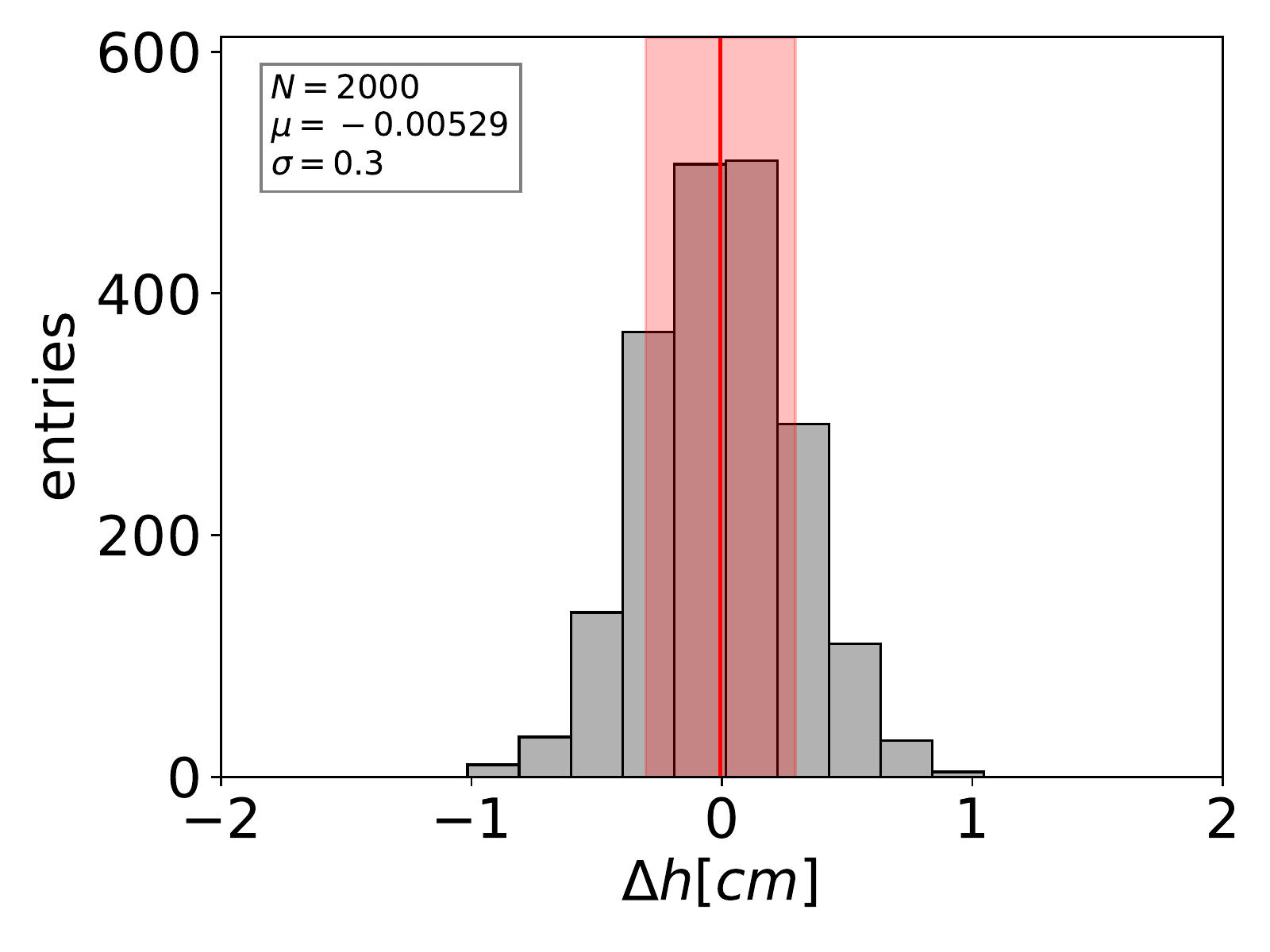}
    \caption{Reconstructed $\Delta h$ using the $\alpha$ and $z_0$ distributions of Fig.~\ref{fig:iceprop_distribution} and a fixed $\Delta n ^{true}$ in Eq.~\eqref{eq:alpha}}
    \label{fig:hreco_distribution}
\end{figure}

Before proceeding to the optimization of the transmitter positions, we show one example for two fixed transmitter positions. Figure \ref{fig:iceprop_distribution} shows the distribution of the reconstructed parameters $\alpha$ (left), $z_0$ (middle) and the correlation $\rho_{\alpha, z_0}$ (right) obtained from 2,000 repetitions. The rms-value of the distribution (red band) is a proxy for the expected precision of the measured ice parameters. For the examples choice of transmitter position (which is the optimal position as we will find later), we find a resolution of $1.6\times 10^{-5}$ and \SI{1}{cm} in $\alpha$ and $z_0$ respectively, as well as a correlation of $1.2\%$. With known $
\Delta n$ we can convert the measurement of $\alpha$ into a measurement of the snow accumulation $\Delta h$ (as discussed earlier). The result is shown in Fig.~\ref{fig:hreco_distribution}. We find a resolution in $\Delta h$ of \SI{3}{mm}.

\subsection{Results}
\label{sec:cal_result}

\begin{figure}[tbp]
\centering
\includegraphics[width=\textwidth]{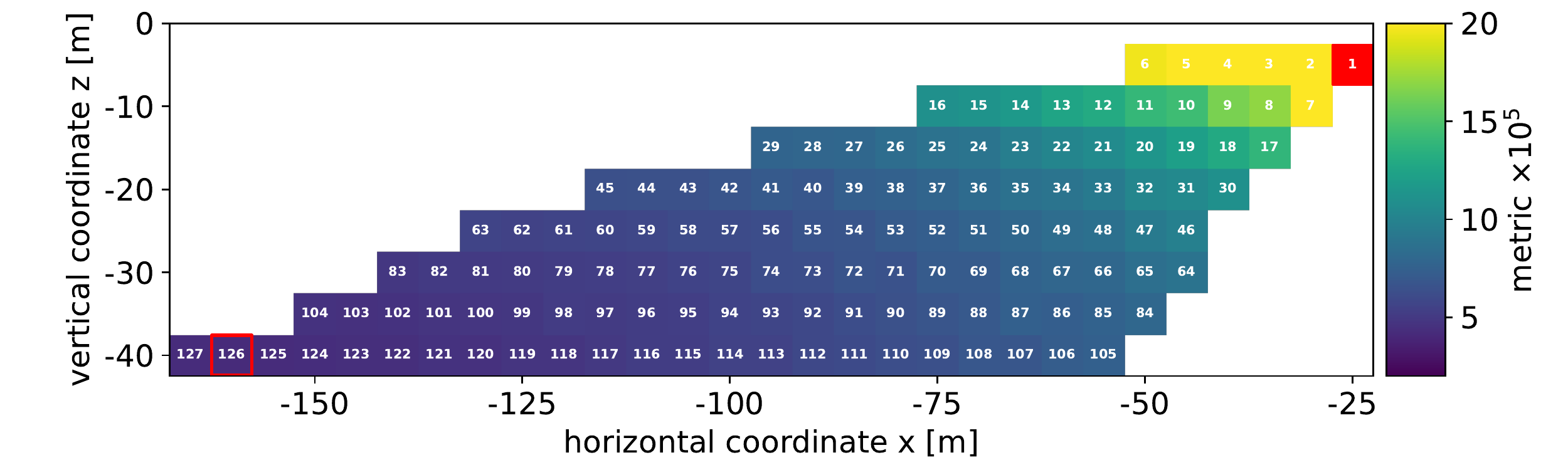}
\includegraphics[width=\textwidth]{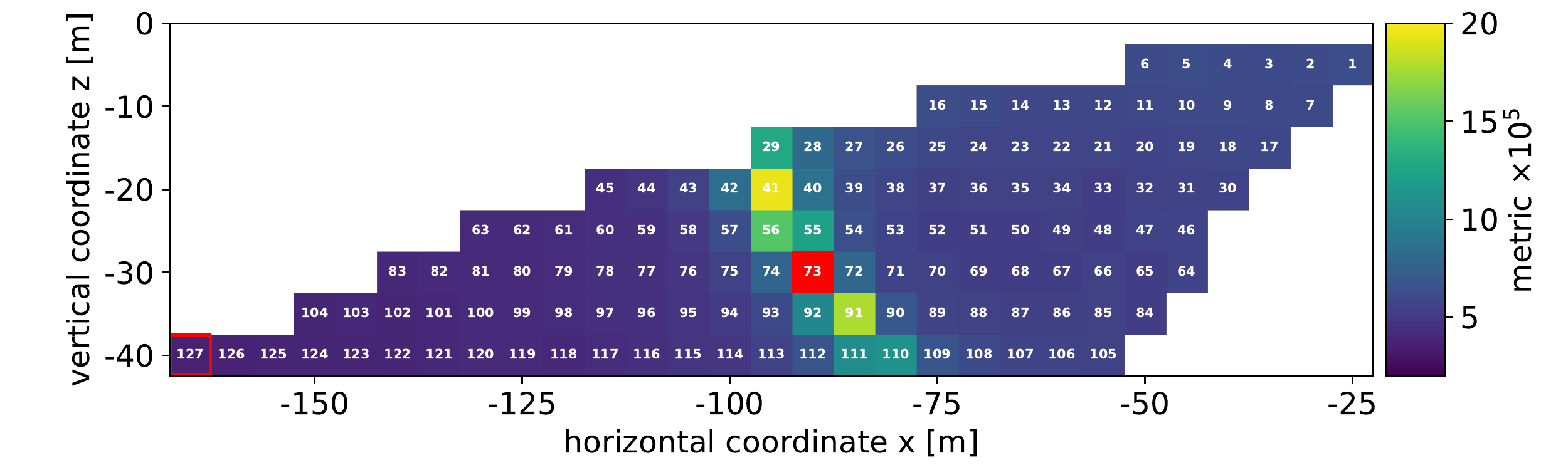}
\caption{Metrics (colour code) as defined by Eq.~\ref{eq:metric} for a fixed transmitter 1 (T1) position and all possible transmitter 2 (T2) positions for the South Pole location. The optimal T1-T2 pair is displayed as the pair of a red-filled pixel (T1) and red-framed pixel (T2). The upper panel shows T1 at position 1, while the lower panel depicts T1 at position 73. Note that the colour bars are scaled by a factor $1\times 10^5$.}
\label{fig:level1}
\end{figure}

Let $N_{pos}$ be all allowed positions on the grid. The optimal transmitter positioning among all possible configurations is then found by first minimizing the metric as defined in Eq.~\ref{eq:metric} of the $(N_{pos}-1)$ T2 positions for a fixed T1 antenna position, which is repeated for all of the $N_{pos}$ possible T1 positions. 
One example is shown in Fig.~\ref{fig:level1} (upper panel) where T1 is placed at the first position (red-filled pixel) on the grid, while the colored pixels indicate the metric for T2 placed at the respective position. The index attributes a number to each position of the grid, starting with 1 in the upper right corner to $N_{pos}=127$ in the lower-left corner. The red-framed pixel marks the position of T2 where the metric is minimal in this particular choice of T1. The plot demonstrates that if the positions of both transmitters are extremely opposed, the best results in the metric are obtained. A smooth gradient from the upper right corner, dominated by large uncertainties, to the lower left can be seen. This can be easily understood when looking at the extreme case, where both transmitters are placed at the same position - the four trajectories would be degenerate since they do not contain any more information than a single transmitter would provide. In Fig.~\ref{fig:level1} (lower panel) we display another configuration where T1 is positioned in the middle of the grid, at index 73. For this case, the optimal position of T2 is at position 127 (red-framed pixel).

\begin{figure}[tbp]
        \centering
        \includegraphics[width=\textwidth]{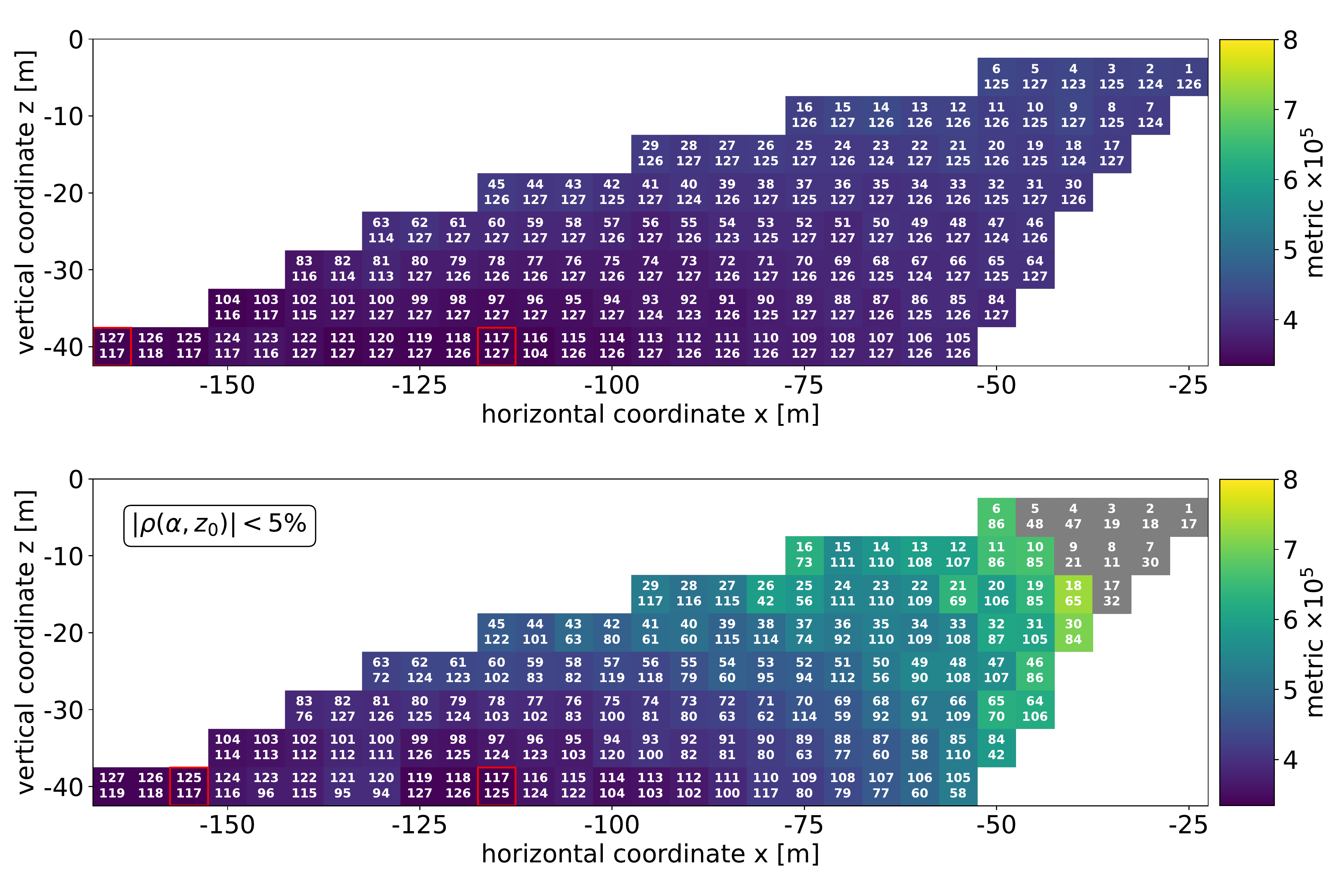}
        \caption{(Upper panel): Metric optimized for pairs of transmitter T1 (upper index in pixel) and T2 (lower index in pixel) at the South Pole. The red-framed pixels indicate the pair for which the smallest value is obtained. (Lower panel): Correlation-constrained, metric-optimized transmitter pairs. The grey pixels in the upper right mark the positions where the metric exceeds a value of $8\times 10^{-5}$, which was set for reasons of visualisation. Note that the colour bars are scaled by a factor $1\times 10^5$.}
        \label{fig:final_sp}
    \end{figure}

In Fig.~\ref{fig:final_sp} (upper panel), we show for each position of T1 (upper index) the optimal position of T2 (lower index). 
There are two configurations, indicated by two red circles, that have the lowest metric of all possible positions. The two circles actually represent the same configuration just with interchanged roles of T1 and T2 ((T1, T2) = (T2, T1)). For a receiver at -15~m at South Pole the optimal transmitter pair is 117-127 corresponding to position T1~=~[-115,~-40]~m and T2~=~[-165,~-40]~m.

Although there is one optimal transmitter pair, we find that there are many pairs of transmitter positions that result in a similarly good metric. Therefore, we also optimize for the transmitter positions that give the least correlation between $\alpha$ and $z_0$. We calculate the correlation via
\begin{equation}
\rho_{\alpha, z_0} = \frac{\sum_{i=1}^{N_{rep}}(\alpha^i - \langle \alpha \rangle)(z_0^i - \langle z_0 \rangle)}{\sqrt{\sum_{i=1}^{N_{rep}}(\alpha^i - \langle \alpha \rangle)^2} \sqrt{\sum_{i=1}^{N_{rep}}(z_0^i - \langle z_0 \rangle)^2}} .
\end{equation}

In a first step, we mask all configurations where the absolute value of the correlation exceeds 5$\%$. Then, for this subset, we determine the transmitter pairs that yield the lowest metric. 
Figure~\ref{fig:final_sp} (lower panel) shows the metric-optimised pairs that fulfill the requirement $|\rho_{\alpha, z_0}|<5\%$. The transmitter pairs differ from the ones in the upper panel of Fig.~\ref{fig:final_sp} and there is a visible gradient in the metric from the upper right to the lower left. The red-framed positions indicate the optimal placement of T1 and T2 amongst the subset of metric optimized and correlation-constrained.

The calculated resolutions depend on the number of measurement points. For now we used 10,000 measurements to increase the precision and robustness of the minimization procedure described above. Although this can be easily obtained by running the calibration system for approximately \SI{15}{minutes} with a repetition rate of the pulser of \SI{10}{Hz}, a smaller time for the calibration run would be preferable to increase the detector uptime. Therefore, we rescale the uncertainties to the expectation for only 100 measurement points. Then, a calibration run would only take 10 seconds and can be run twice a day without introducing relevant downtime for neutrino detection.  
The per-100-events parameter uncertainty is a factor of 10 ($\sqrt{10,000/100}$) higher than what the plots suggest, which we have verified for some configurations. The correlation is not affected by a reduced number of measurements. Table~\ref{tab:southpole} list the metric, the correlation as well as the per-100-events rms-value for the distribution in $\alpha$ and $z_0$ of the metric-optimised pair (upper part, c.f Fig.~\ref{fig:final_sp} (upper panel)) and the correlation-constrained, metric-optimised pair (lower part, c.f. Fig.~\ref{fig:final_sp} (lower panel)).
As can be seen from Tab.~\ref{tab:southpole}, by constraining the allowed transmitter configurations to those with low correlation, we are able to reduce $\rho_{\alpha, z_0}$ from $14\%$ to $1.2\%$, while not considerably increasing the metric or the parameter uncertainty. We find a relative uncertainty in $\alpha$ and $z_0$ of $0.04\%$ and $0.14\%$ respectively. 

    \begin{table}[tbp]
        \centering
        \smallskip
        \begin{tabular}{cc|cccc}
            \hline\hline
            \multicolumn{6}{c}{metric-optimised configuration}\\
            \hline
            T1 & T2 & $\mathcal{M}$ & $\rho_{\alpha, z_0}$ & $\sigma_{\alpha}/\alpha^\mathrm{true}$ & $\sigma_{z_0}/z_0^\mathrm{true}$ \\
            $[-115,\ -40]$~m & $[165,\ -40]$~m & $3.4\cdot 10^{-5}$ & 14$\%$ & 0.03$\%$ & 0.14$\%$\\ 
            \hline\hline
            \multicolumn{6}{c}{|correlation| < 5$\%$ + metric-optimised configuration}\\
            \hline
            T1 & T2 & $\mathcal{M}$ & $\rho_{\alpha, z_0}$ & $\sigma_{\alpha}/\alpha^\mathrm{true}$ & $\sigma_{z_0}/z_0^\mathrm{true}$ \\
            $[-115,\ -40]$~m & $[155,\ -40]$~m & $3.4\cdot 10^{-5}$ & 1.2$\%$ & 0.04$\%$ & 0.14$\%$\\ 
            \hline\hline
        \end{tabular}
        \caption{\label{tab:southpole} Optimal pair and corresponding metric, correlation and per-100-events parameter uncertainty at South Pole for the metric-optimised pair (configuration as in Fig.~\ref{fig:final_sp} (upper panel)) in the upper part and correlation-constrained, metric-optimised pair (configuration as in Fig.~\ref{fig:final_sp} (lower panel)) in the lower part. We have flipped the $x$ coordinate of T2 and placed it in the lower right quadrant as depicted in Fig.~\ref{fig:southpole}.}
    \end{table}

In summary, we find that for the South Pole site, a calibration system comprised of two transmitter antennas positioned at T1~=~[-115,~-40]~m and T2~=~[-155,~-40]~m yields optimal results. As the setup is symmetrical under translation in $x$, we can place T2 into the right quadrant as depicted in Fig.~\ref{fig:southpole} which additionally allows probing a potential tilt in snow accumulation. In the following, we therefore quote the position of T2 to have a positive $x$ coordinate.

We also studied allowing deeper deployment depth. The rationale is that a hybrid detector station, such as used in the RNO-G detector, anyway have holes drilled up to \SI{200}{m}. Then, we find  T1~=~[-310,~-200]~m and T2~=~[390,~-200]~m as the optimal antenna positions, yielding a resolution of $0.02\%$ in $\alpha$ and $0.07\%$ in $z_0$ and a correlation of 4.7$\%$. The resolution improves as expected because of the larger path lengths that reduce the impact of the fixed timing uncertainty (of \SI{0.2}{ns} in this study). However, we also found that deeper transmitter also need to be placed further away horizontally to have a large reflection coefficient at the surface, as already visible from Fig.~\ref{fig:grid_sp}. The transmitter needs to be at least as far away horizontally as deep. Hence, the existing deep instrumentation holes of a hybrid detector station can not be used to install the calibration transmitters as their horizontal extend is only $\mathcal{O}$(\SI{20}{m}). Thus, the deployment of the calibration transmitters to deeper depths would require drilling of new holes which is expensive and therefore disfavors doing it as the benefit of going deeper is small for the purposes of measuring the snow accumulation and the n(z) profile. 

We also repeated the study with a more restricted deployment depth of \SI{20}{m} to speed up the deployment time (the drill time increases more than linear with depth).  Then, the optimal transmitter positions are T1~=~[-85,~-20]~m and T2~=~[115,~-20]~m with a resulting uncertainty of $0.05\%$ and $0.21\%$ in $\alpha$ and $z_0$ and a correlation of 3.2$\%$. The conclusion that can be drawn from that is, that the optimal positions of the antennas tend to lay as deep as possible, while the parameter uncertainty reduces a factor of 2 when increasing the depth by a factor of 10. We consider restricting the deployment depth to \SI{20}{m} a viable option if required by deployment constraints. 
    
\subsection{Systematic Uncertainties}
\label{sec:cal_syst}
We demonstrated that a calibration setup with a per-100-event statistical uncertainty of 0.04$\%$ in $\alpha$ and 0.14$\%$ in $z_0$ is achievable. Additionally, if $\alpha$ and $z_0$ are measured and $\Delta n$ is known, $\Delta h$ can be reconstructed with a resolution of \SI{3}{mm}. In the following, two sources of systematic uncertainties are discussed: Detector geometry uncertainty (Sec.~\ref{sec:cal_syst_geo}) and surface flatness (Sec.~\ref{sec:cal_syst_fresnel}).

\subsubsection{Detector Geometry Uncertainty}
\label{sec:cal_syst_geo}
The detector geometry constitutes a systematic error to the analysis due to the limited deployment accuracy, affecting both the horizontal and vertical location of the antenna. So far the analysis did not include any uncertainties regarding the spacial position of the transmitter and receiver antenna. Experience from the ARIANNA experiment has shown that a deployment precision of $\pm$\SI{10}{mm} in $x$ and $z$ is feasible. An uncertainty of $\pm$\SI{10}{mm} in $z$ directly propagates to an equal systematic error in the absolute snow accumulation. The relative snow accumulation, that is the change in snow height, is unaffected by that. 

We study the impact of a transmitter deployment uncertainty on the reconstruction of $\alpha$ and $z_0$ for the optimal transmitter configurations found in Tab.~\ref{tab:southpole} with a fixed snow accumulation $\Delta h$=\SI{0}{cm}. First, we do the ray-tracing with the unperturbed ice properties but with slightly displaced antennas. Second, we fit the parameters $\alpha$ and $z_0$ assuming the unperturbed antenna geometry using the same minimization procedure as described above. This will result in slightly modulated ice properties. In this study four displacements in the horizontal $\Delta x$ and vertical $\Delta z$ direction plus the ideal, non-displaced case are considered which can be written in tuples $(\Delta x,\Delta z)$: (0,0)~cm, (1,0)~cm, (-1,0)~cm, (0,1)~cm and (0,-1)~cm. That accounts for $5 \times 5 = 25$ distinct displacements (1 unperturbed + 24 perturbed). The positional deviations are added to the position of R and T1. T2 remains untouched as only the relative positions between the antennas are relevant here. An uncertainty in the absolute position of T2 is already captured in the relative uncertainties of the positions of R and T1 relative to T2. For each of the 25 displacement pairs, we fit 500 different random realizations of the observables. 

\begin{figure}[t]
    \centering
    \includegraphics[width=\textwidth]{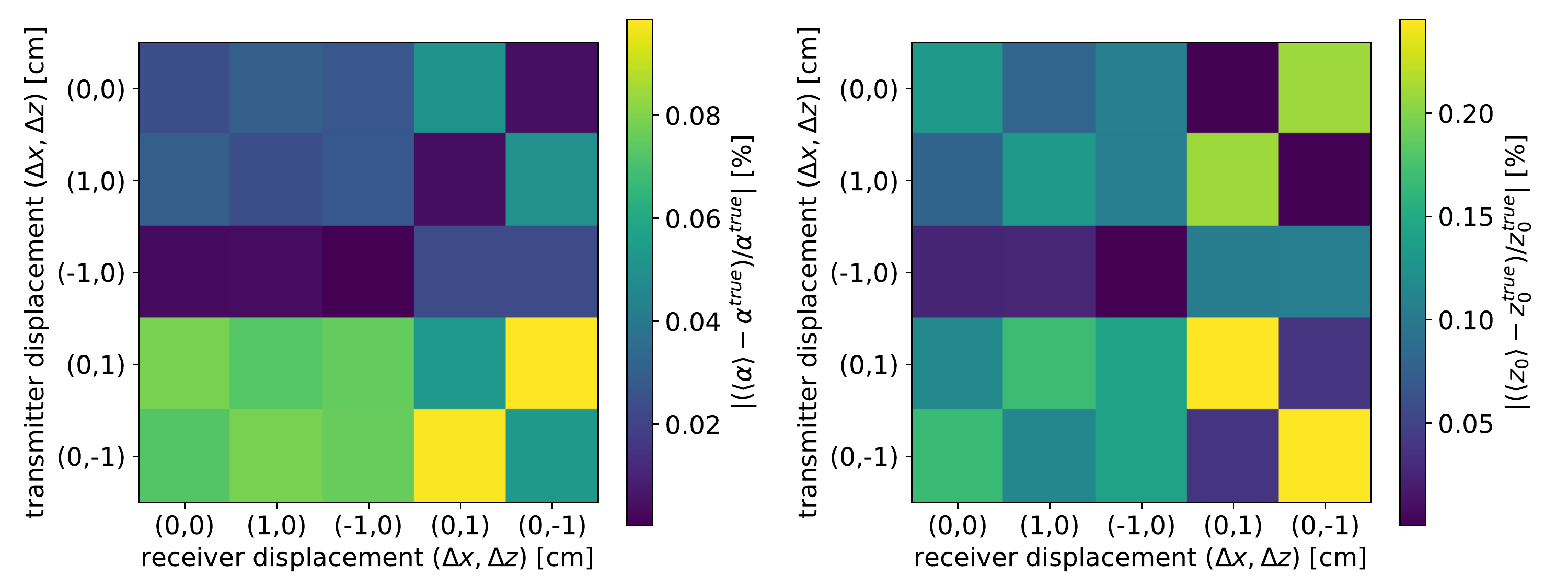}
    \caption{The absolute value of the relative deviation in $\alpha$ (left) and $z_0$ (right) for different combinations of receiver and transmitter displacements ($\Delta x,\ \Delta z$) for the South Pole.}
    \label{fig:cal_syst_geo}
\end{figure}

Figure~\ref{fig:cal_syst_geo} gives a visual impression of the reconstruction capabilities in $\alpha$ (left) and $z_0$ (right) for all considered combinations of displacements $(\Delta x,~\Delta z)$ in R and T1. The color code shows the absolute value of the relative deviation between the median of the perturbed distribution to the unperturbed distribution. The average over all 24 perturbed placements gives a systematic uncertainty of $0.04\%$ in $\alpha$ and $0.11\%$ in $z_0$ compared to a statistical error of $0.04\%$ in $\alpha$ and $0.14\%$ in $z_0$ from Tab.~\ref{tab:southpole}. We find that with the current setup, systematic and statistical uncertainties are similar. 

Results from simulations performed with a less optimistic antenna displacement of \SI{10}{cm} and \SI{100}{cm} suggest that the systematic uncertainty in $\alpha$ and $z_0$ scales linearly. For \SI{100}{cm} antenna displacement for instance we find an averaged, systematic error of 4.4$\%$ in $\alpha$ and 10.9$\%$ in $z_0$, which is a factor of 100 larger than for a \SI{1}{cm} displacement. We later find that even with 10-fold increase of the antenna positioning uncertainty of \SI{10}{cm} we still get acceptable results (c.f.~\ref{sec:nuprop_iceprop}).

\subsubsection{Surface Flatness}
\label{sec:cal_syst_fresnel}
The drawing shown in Fig.~\ref{fig:southpole} seems to indicate that the reflection takes place at a single point. However, the reflection takes place over an extended region described by the Fresnel zone. An estimate of the spatial size of the first Fresnel zone $F_1$ (Eq.~\eqref{eq:fresnel1}) at distance $d_t$ ($d_r$) from the transmitter (receiver) antenna for signals of wavelength $\lambda$ is given by

\begin{align}
F_1 = \sqrt{\frac{\lambda d_t d_r}{d_t+d_r}}.
\label{eq:fresnel1}
\end{align}

For emitter position T1 at South Pole we approximately have $d_t$ = \SI{79}{m} and $d_r$ = \SI{36}{m} which yields for a typical frequency of \SI{200}{MHz} a size of the Fresnel zone of \SI{6}{m}. Similarly, for transmitter position T2 we get an approximate size of \SI{7}{m}. Hence, the calibration system already probes the average snow accumulation over an extended region and averages over small-scale fluctuations. However, of relevance for neutrino detection is the complete area around the detector station where neutrino signals could be reflected. Therefore, any local variations of snow height need to be studied and if relevant taken into account.

For a neutrino-induced radio signal, the Fresnel zone is of similar size. The distance from the surface reflection area to the ``emitter'' (i.e. the neutrino interaction in the ice) is much larger than the path $d_r$ from the surface down to the receiving antenna. Then Eq.~\ref{eq:fresnel1} simplifies to
\begin{align}
F_1 = \sqrt{\lambda d_r} =  \sqrt{\lambda \frac{\SI{15}{m}}{\sin(\theta)}} ,
\label{eq:fresnel2}
\end{align}
where \SI{15}{m} is the depth of the receiving antenna. Now, the size of the Fresnel zone only depends on the signal arrival direction which typically varies between $\theta$ = \SI{10}{\degree} to \SI{30}{\degree} below the horizon \cite{DnR2019} which results in sizes of the Fresnel zones of \SI{7}{m} to \SI{11}{m}. The corresponding centers of the Fresnel zones are \SI{26}{m} up to \SI{85}{m} away from the station center. Thus, a large part of the area around the detector station is not probed explicitly by the calibration setup. As the site can be easily surveyed during the installation of the station, we only discuss scenarios where a flat surface would develop local irregularities over time. Two scenarios are conceivable that we discuss in the following:

First, sastrugi, i.e., elongated snow hills, that are formed and drift due to katabatic winds would lead to local differences in snow height, typically ranging from \SI{1}{m} - \SI{2}{m} in length and \SI{10}{cm} - \SI{15}{cm} in height \cite{Sastrugi1965, Sastrugi1998}. Because their typical lateral extent is smaller than the Fresnel zone, typical sastrugi have likely no significant effect on the reflection measurement and the average height over the Fresnel zone is what matters. However, also larger sastrugi are possible which we would be able to probe with the calibration system. The expected experimental signature of large-enough sastrugi is to first see an increase in snow accumulation when a sastruga is formed within one of the two Fresnel zones that are probed by the D'n'R measurement. Over time the peak of the sastruga would wash out leading to a decrease in the measured snow accumulation. During the next storm, new sastrugi would get formed at different positions leading to different measured snow accumulations at the two Fresnel zones that are being probed. If a measurement at the South Pole would show consistent snow accumulations at both Fresnel zones and no decrease with time, it would disfavor the presence of sastrugi that are large enough to impact the radio measurements. 
At Moore's Bay, Antarctica, where the prototype of this calibration setup operated for a year, no decrease in snow accumulation was observed. The observation was that the snow accumulation jumped up by a few centimeters during a storm, and then stayed constant until the next increase \cite{DnR2019,ICRC21_calibration}. However, more measurements are needed to draw firm conclusions about the presence of sastrugi. 

Second, a global tilt of the snow surface, or a more complex but smooth location-dependent change in snow height could form over time. To probe this effect we suggest placing the transmitters in opposite directions from the receiving antennas as indicated in Fig.~\ref{fig:southpole}. Then, the calibration setup already probes two positions roughly \SI{87}{m} apart. If a large-scale tilt would develop over time, it should show up as a difference in snow accumulation between the two positions. If both positions show the same snow accumulation over time, it would disfavor this scenario.

\section{Impact on the Reconstruction of Neutrino Properties}
\label{sec:nuprop}

In this section, we study the impact of uncertainties in the index-of-refraction profile $n(z)$ and the snow accumulation on the reconstruction of the neutrino vertex distance, direction, and energy for the South Pole site. The goal of this section is to find out how well the snow accumulation and the $n(z)$ profile needs to be known to not impact the ability to measure the neutrino properties of interest. 
Section~\ref{sec:nuprop_sim} describes the general simulation setup, while in Sec.~\ref{sec:nuprop_iceprop} and \ref{sec:nuprop_snowheight} we present the results separated for variations in the ice properties and snow height respectively. We only study the impact of uncertainties in the $n(z)$ profile and snow accumulation. We do not consider other sources of uncertainties on the reconstruction of the neutrino vertex distance, direction, and energy that are likely dominant. 

\subsection{Simulation Setup}
\label{sec:nuprop_sim}

We use the same MC data set as used in \cite{DnR2019} to determine the obtainable neutrino vertex distance resolution with the D'n'R technique: A detailed NuRadioMC simulation \cite{NuRadioMC2019,NuRadioReco2019} that includes the simulation of the initial neutrino interaction, followed by radio signal generation and propagation to a detailed detector simulation.
For simulated neutrino energies of $10^{17}$~eV ($10^{18}$~eV) the data set contains 1431 (9463) events that pass the trigger threshold. The simulation output that is relevant for the following study is the distribution of the neutrino interaction vertices of the triggered events. The larger the neutrino energy, the further away is the neutrino interaction vertex on average.

We study the impact of deviations in the index-of-refraction profile $n(z)$ and the snow accumulation on the neutrino vertex distance $R$, the neutrino direction and the neutrino energy $E_\nu$ reconstruction. 
We quantify the neutrino direction reconstruction as the space angle between the nominal and the reconstructed launch vector of the direct signal. 
The neutrino vertex distance $R$ is given by the propagation length of the radio signal from its generation to the receiving antenna.
The distance $R$ can be determined from the time difference between the direction and reflected signal trajectory to a approx.~\SI{15}{m} deep Vpol antenna. We use the lookup table from \cite{DnR2019} that translates the D'n'R time difference $\Delta T$ and the zenith angle of the signal arrival direction to the vertex distance $R$. 
The uncertainty in vertex distance is itself not very meaningful but what matters is how this uncertainty impacts the estimate of the neutrino energy which requires correcting the measured signal for the signal attenuation due to the propagation through the ice.
Following the same procedure as in \cite{DnR2019}, we calculate the impact of the distance uncertainty on the neutrino energy by  calculating a ``pseudo'' shower energy $E$ as given in Eq.~\eqref{eq:shower_energy}, with $L_\mathrm{att}$ being the attenuation length in ice. The attenuation length is depth and frequency dependent which would need to be taken into account when reconstructing measured events. In our estimate here, we use an average value of \SI{1}{km} \cite{Barwick2005} which is representative for most geometries. 

\begin{align}
    E \propto \frac{R}{e^{R/L_\mathrm{att}}}
    \label{eq:shower_energy}
\end{align}
This formula is essentially correcting a unit measured signal for attenuation. We note that a measurement of the neutrino energy requires additional steps, such as the reconstruction of the electric field and a measurement of the viewing angle, but as these measurements are not impacted by uncertainties in snow accumulation or the $n(z)$ profile, we ignore them here. For a comprehensive description of reconstruction strategies we refer the reader to \cite{BarwickGlaser2022}.

We will first focus on variations in the index-of-refraction profile $n(z)$ (Sec.~\ref{sec:nuprop_iceprop}) within the uncertainties of the combined statistical and systematic uncertainty found in Sec.~\ref{sec:cal_result} and \ref{sec:cal_syst_geo}. We consider deviations in $\alpha$ and $z_0$ of $[- \sigma_{i},~+ \sigma_{i}]$, with $\sigma_{i} = \sqrt{(\sigma_{i}^{\mathrm{stat}})^2+(\sigma_{i}^{\mathrm{syst}})^2}$ and $i = \Delta n,~z_0$. In total we simulate four combinations of deviations in $\Delta n$ and $z_0$ which result in four pairs $(\delta \Delta n, \delta z_0)\ =\ (-1,-1)\sigma,\ (-1,+1)\sigma,\ (+1,-1)\sigma,\ (+1,+1)\sigma$.

In addition, we study how an inaccurate measurement of the snow accumulation (Sec.~\ref{sec:nuprop_snowheight}), either during deployment or due to local variations of the snow level, affects the reconstruction of neutrino properties. We consider a wide range of snow height deviations from \SI{1}{cm} to \SI{100}{cm}.

For each of the simulated events we run the ray-tracing algorithm for the unperturbed case which yields the nominal vertex distance $R_\mathrm{nom}$ and launch vector of the direct radio pulse $\vec{v}_\mathrm{nom}$ as well as the proxy shower energy $E_\mathrm{nom}$. Subsequently, the perturbed model is applied and the reconstructed properties $R_\mathrm{reco}$, $\vec{v}_\mathrm{reco}$ and $E_\mathrm{reco}$ are computed. We then define the deviation in the neutrino direction reconstruction as $\measuredangle{(\vec{v}_\mathrm{nom}, \vec{v}_\mathrm{reco})}$.

Ideally, we would want to keep the additional uncertainty in the reconstructed energy introduced by variations of the ice properties smaller than the natural limit of about a factor of two due to the unknown inelasticity, i.e., the amount of neutrino energy transferred into the particle shower \cite{DnR2019}. If that is the case we can set limits on the acceptable precision of those parameters that do not worsen the neutrino energy reconstruction by a factor of 2 and direction by less than a degree (the expected angular resolution of future  shallow radio detectors is 1 to a few degrees depending on event quality, see e.g. Ref.~\cite{Hallmann2021,Glaser:2019kjh,ARIANNA:2021pzm}).

\subsection{Variations of the Ice Properties}
\label{sec:nuprop_iceprop}

\begin{figure}[t]
    \centering
    \includegraphics[width=\textwidth]{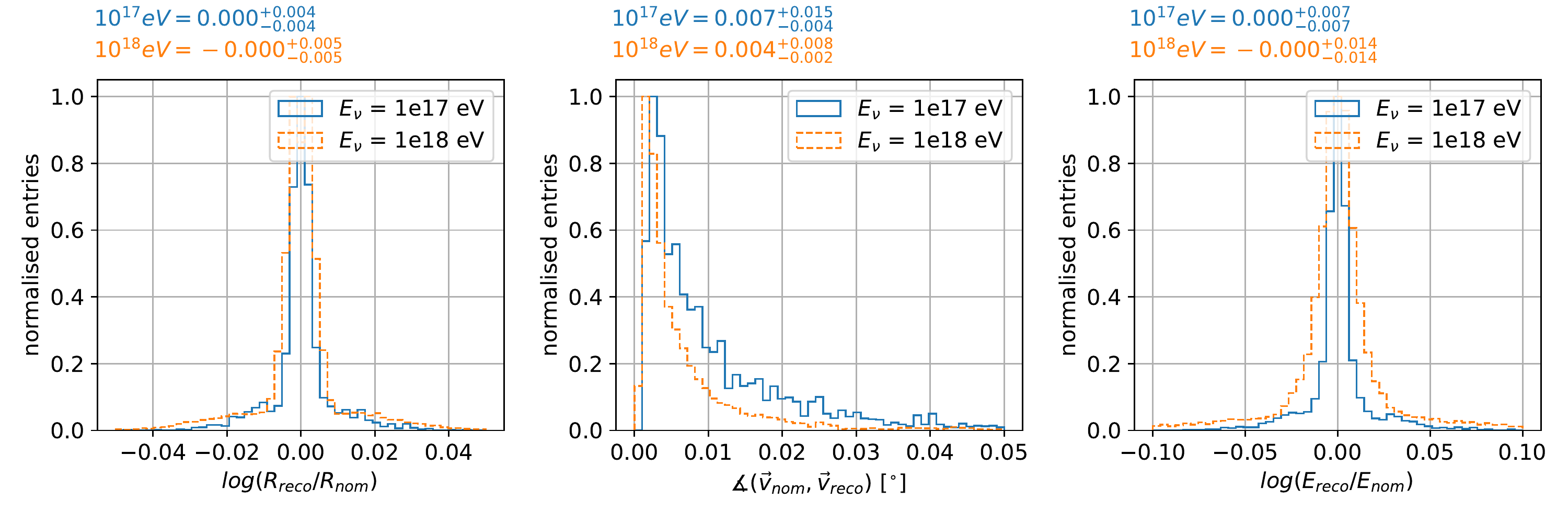}
    \caption{Distribution of nominal and reconstructed neutrino vertex distance $R$ (left), neutrino direction (middle) the and neutrino energy $E$ (right) for energies of $10^{17}$~eV (blue) and $10^{18}$~eV (orange) for a $1\sigma$ deviation in $\Delta n $ and $z_0$. The distance is measured in units of length (e.g. meters) and the energy in electron volts. Here, we only show the relative deviation to the nominal (or true) distance/energy which simplifies parameterizing the distributions. The colored text above the figure indicates the median and the 68\% quantiles of the distribution.}
    \label{fig:nuprop_icemodel}
\end{figure}

We simulated a combined relative uncertainty in $\sigma_\mathrm{comb}(\alpha) = 0.07\%$ and $\sigma_\mathrm{comb}(z_0) = 0.19\%$ for both $10^{17}$~eV and $10^{18}$~eV neutrino events. For each of the four perturbation pairs we obtain a distribution in $\log_{10}(R_\mathrm{reco}/R_\mathrm{nom})$, $\measuredangle{(\vec{v}_\mathrm{nom}, \vec{v}_\mathrm{reco})}$ and $\log_{10}(E_\mathrm{reco}/E_\mathrm{nom})$. Figure~\ref{fig:nuprop_icemodel} depicts the combination of the four perturbations in one histogram for vertex distance $R$ (left), direction (middle) and energy $E$ (right), while the colored text above the figure notes the median and the 68$\%$ quantile of the distribution. We find a 68$\%$ quantile for $10^{17}$~eV ($10^{18}$~eV) in $\log_{10}(R_\mathrm{reco}/R_\mathrm{nom})$ of 0.004 (0.005) which translates to 0.9$\%$ (1.2$\%$) on a linear scale. For $\log_{10}(E_\mathrm{reco}/E_\mathrm{nom})$ we find 0.007 (0.014) which translates to 1.6$\%$ (3.3$\%$) on a linear scale. 

For comparison, the statistical uncertainty of the measured D'n'R time delay $\Delta t$ and signal arrival direction results in an uncertainty of $\log_{10}(E_\mathrm{reco}/E_\mathrm{nom})$ of 0.08 (0.15) at $10^{17}$~eV ($10^{18}$~eV) \cite{DnR2019}. Similarly, an uncertainty of $\log_{10}(R_\mathrm{reco}/R_\mathrm{nom})$ of 0.04 (0.05) at $10^{17}$~eV ($10^{18}$~eV) was found \cite{DnR2019}. Thus, the calibration system described above provides a measurement of the $n(z)$ profile where the systematic uncertainties from the $n(z)$ profile are smaller than the statistical uncertainties from the time delay obtained from the DnR measurement. Therefore, there is no strong reason to further improve the measurement of $n(z)$.  

\begin{table}[tbp]
    \centering
    \begin{tabular}{ccccccc}
    \hline
    \hline
    & \multicolumn{2}{c}{$\log_{10}(R_\mathrm{reco}/R_\mathrm{nom})$} & \multicolumn{2}{c}{$\measuredangle{(\vec{v}_\mathrm{nom}, \vec{v}_\mathrm{reco})}\ [^{\circ}]$} & \multicolumn{2}{c}{$\log_{10}(E_\mathrm{reco}/E_\mathrm{nom})$}\\
    \hline
    $\# \sigma$ & $10^{17}$~eV & $10^{18}$~eV & $10^{17}$~eV & $10^{18}$~eV & $10^{17}$~eV & $10^{18}$~eV \\
    \hline
    1 & $-0.000^{+0.004}_{-0.004}$ & $-0.000^{+0.005}_{-0.005}$ & $0.007^{+0.015}_{-0.004}$ & $0.004^{+0.008}_{-0.002}$ & $0.000^{+0.0007}_{-0.0007}$ & $-0.000^{+0.014}_{-0.014}$\\
    10 & $-0.001^{+0.037}_{-0.035}$ & $-0.003^{+0.054}_{-0.048}$ & $0.070^{+0.142}_{-0.043}$ & $0.038^{+0.077}_{-0.021}$ & $-0.001^{+0.071}_{-0.066}$ & $-0.005^{+0.145}_{-0.129}$\\
    100 & $-0.096^{+0.247}_{-0.293}$ & $-0.224^{+0.346}_{-0.310}$ & $0.634^{+1.113}_{-0.373}$ & $0.339^{+0.517}_{-0.178}$ & $-0.159^{+0.441}_{-0.498}$ & $-0.497^{+0.815}_{-0.609}$\\
    \hline
    \hline
    \end{tabular}
    \caption{\label{tab:nuprop_icemodel} Reconstruction of the neutrino vertex distance $R$, the neutrino direction and the neutrino energy $E$ for energies of $10^{17}$~eV and $10^{18}$~eV and $1\sigma$, $10 \sigma$ and $100 \sigma$ deviations in $\Delta n $ and $z_0$.}
\end{table}

In an additional crosscheck, we studied unrealistic deviations in $\Delta n$ and $z_0$ of 10 and 100 times the above uncertainty. These results are listed in Tab.~\ref{tab:nuprop_icemodel}. We conclude that even with uncertainties in the ice properties of more than a factor of 10 larger than our nominal reconstruction capabilities, the energy resolution does not worsen with respect to the constraints set by the inelasticity limit. 

\subsection{Variations of the Snow Height}
\label{sec:nuprop_snowheight}

We consider a systematic uncertainty in the snow height, that is the difference between the perturbed snow height $\Delta h_\mathrm{reco}$ and the nominal snow height $\Delta h_\mathrm{nom}$ of $\delta \Delta h = \Delta h_\mathrm{nom} - \Delta h_\mathrm{reco}$ with $\delta \Delta h$ ranging from \SI{+1}{cm} to \SI{+100}{cm}. Thus, a positive value for $\delta \Delta h$ has the same effect as ``burying'' the antennas (and neutrino vertices) deeper into the ice. Table~\ref{tab:nuprop_snowheight} lists the considered perturbations and as well as the median and the 68$\%$ quantiles of the so obtained distributions in $\log_{10}(R_\mathrm{reco}/R_\mathrm{nom})$, $\measuredangle{(\vec{v}_\mathrm{nom}, \vec{v}_\mathrm{reco})}$ and $\log_{10}(E_\mathrm{reco}/E_\mathrm{nom})$. Generally, the uncertainty is higher for higher energies. We find that, for small scale variations in the snow height of \SI{+1}{cm} the effect on the neutrino reconstruction parameters are negligible. For a deviation of \SI{+10}{cm}, the resulting uncertainty in reconstructed energy is similar to the expected statistical uncertainty but still below the intrinsic uncertainty from inelasticity fluctuations of 0.3 in $\log_{10}(E)$. Thus, even an uncertainty of \SI{20}{cm} in snow height seems tolerable. 
The resulting uncertainty in the neutrino direction is negligible in all cases. We further simulated negative deviations between \SI{-1}{cm} to \SI{-100}{cm}, i.e. effectively lifting the antennas (and neutrino vertices). The scale of the effect is comparable to the above but the distributions in $\log_{10}(R_\mathrm{reco}/R_\mathrm{nom})$ and $\log_{10}(E_\mathrm{reco}/E_\mathrm{nom})$ are mirrored, with one distinction being that for high deviations of \SI{\pm100}{cm} uncertainties for $10^{17}$~eV neutrinos are larger than for $10^{18}$~eV neutrinos. 

\begin{table}[tbp]
    \centering
    \begin{tabular}{ccccccc}
    \hline
    \hline
    & \multicolumn{2}{c}{$\log_{10}(R_\mathrm{reco}/R_\mathrm{nom})$} & \multicolumn{2}{c}{$\measuredangle{(\vec{v}_\mathrm{nom}, \vec{v}_\mathrm{reco})}\ [^{\circ}]$} & \multicolumn{2}{c}{$\log_{10}(E_\mathrm{reco}/E_\mathrm{nom})$}\\
    \hline
    $\delta \Delta h$ [cm] & $10^{17}$~eV & $10^{18}$~eV & $10^{17}$~eV & $10^{18}$~eV & $10^{17}$~eV & $10^{18}$~eV \\
    \hline
    +1 & $-0.00^{+0.00}_{-0.00}$ & $-0.00^{+0.00}_{-0.01}$ & $0.00^{+0.00}_{-0.00}$ & $0.00^{+0.00}_{-0.00}$ & $-0.01^{+0.01}_{-0.01}$ & $-0.01^{+0.01}_{-0.02}$\\
    +2 & $-0.00^{+0.00}_{-0.01}$ & $-0.01^{+0.01}_{-0.02}$ & $0.00^{+0.00}_{-0.00}$ & $0.00^{+0.00}_{-0.00}$ & $-0.01^{+0.01}_{-0.02}$ & $-0.01^{+0.02}_{-0.05}$\\
    +5 & $-0.01^{+0.01}_{-0.03}$ & $-0.01^{+0.01}_{-0.04}$ & $0.00^{+0.01}_{-0.00}$ & $0.00^{+0.00}_{-0.00}$ & $-0.02^{+0.01}_{-0.06}$ & $-0.03^{+0.03}_{-0.11}$\\
    +10 & $-0.02^{+0.02}_{-0.05}$ & $-0.03^{+0.02}_{-0.07}$ & $0.01^{+0.01}_{-0.00}$ & $0.00^{+0.01}_{-0.00}$ & $-0.03^{+0.03}_{-0.11}$ & $-0.06^{+0.06}_{-0.19}$\\
    +20 & $-0.04^{+0.03}_{-0.10}$ & $-0.05^{+0.05}_{-0.13}$ & $0.01^{+0.03}_{-0.01}$ & $0.01^{+0.01}_{-0.00}$ & $-0.07^{+0.06}_{-0.19}$ & $-0.12^{+0.11}_{-0.32}$\\
    +50 & $-0.10^{+0.07}_{-0.21}$ & $-0.13^{+0.12}_{-0.26}$ & $0.03^{+0.06}_{-0.02}$ & $0.02^{+0.03}_{-0.01}$ & $-0.17^{+0.13}_{-0.37}$ & $-0.28^{+0.25}_{-0.56}$\\
    +100 & $-0.19^{+0.14}_{-0.34}$ & $-0.24^{+0.23}_{-0.39}$ & $0.06^{+0.13}_{-0.04}$ & $0.03^{+0.07}_{-0.02}$ & $-0.32^{+0.23}_{-0.54}$ & $-0.50^{+0.45}_{-0.72}$\\
    \hline
    -1 & $0.00^{+0.00}_{-0.00}$ & $0.00^{+0.01}_{-0.00}$ & $0.00^{+0.00}_{-0.00}$ & $0.00^{+0.00}_{-0.00}$ & $0.01^{+0.01}_{-0.01}$ & $0.01^{+0.02}_{-0.01}$\\
    -2 & $0.00^{+0.01}_{-0.00}$ & $0.01^{+0.02}_{-0.01}$ & $0.00^{+0.00}_{-0.00}$ & $0.00^{+0.00}_{-0.00}$ & $0.01^{+0.03}_{-0.01}$ & $0.01^{+0.05}_{-0.02}$\\
    -5 & $0.01^{+0.03}_{-0.01}$ & $0.01^{+0.04}_{-0.01}$ & $0.00^{+0.01}_{-0.00}$ & $0.00^{+0.00}_{-0.00}$ & $0.02^{+0.07}_{-0.02}$ & $0.03^{+0.13}_{-0.03}$\\
    -10 & $0.02^{+0.06}_{-0.02}$ & $0.02^{+0.08}_{-0.02}$ & $0.01^{+0.01}_{-0.00}$ & $0.00^{+0.01}_{-0.00}$ & $0.04^{+0.14}_{-0.03}$ & $0.06^{+0.25}_{-0.05}$\\
    -20 & $0.04^{+0.12}_{-0.03}$ & $0.04^{+0.12}_{-0.03}$ & $0.01^{+0.03}_{-0.01}$ & $0.01^{+0.01}_{-0.00}$ & $0.07^{+0.28}_{-0.05}$ & $0.09^{+0.37}_{-0.08}$\\
    -50 & $0.07^{+0.19}_{-0.05}$ & $0.05^{+0.16}_{-0.05}$ & $0.03^{+0.06}_{-0.02}$ & $0.02^{+0.03}_{-0.01}$ & $0.12^{+0.47}_{-0.09}$ & $0.13^{+0.49}_{-0.11}$\\
    -100 & $0.11^{+0.23}_{-0.08}$ & $0.07^{+0.17}_{-0.06}$ & $0.06^{+0.12}_{-0.04}$ & $0.03^{+0.07}_{-0.02}$ & $0.18^{+0.59}_{-0.13}$ & $0.14^{+0.50}_{-0.12}$\\
    \hline
    \hline
    \end{tabular}
    \caption{\label{tab:nuprop_snowheight} Reconstruction of the neutrino vertex distance $R$, the neutrino direction and the neutrino energy $E$ for energies of $10^{17}$~eV and $10^{18}$~eV and deviations in $\Delta h$ between \SI{1}{cm} and \SI{100}{cm} (upper half) and between \SI{-1}{cm} and \SI{-100}{cm} (lower half).}
\end{table}

\section{Conclusion}
\label{sec:conclusion}

The D’n’R technique provides unique opportunities for the detection of high-energy neutrinos via the Askaryan effect. An antenna placed $\mathcal{O}$(\SI{15}{m}) below the ice surface will measure a direct pulse and a pulse reflected off the ice surface for most detected neutrino interactions in the ice. This signature not only provides a unique characteristic of a neutrino origin of the signal but also allows to measure the distance to the neutrino interaction vertex precisely, a crucial property to determine the neutrino energy and direction. 
The D’n’R technique was tested experimentally at Moore’s Bay on the Ross ice shelf and used for continuous monitoring of the snow accumulation which is proportional to the time delay between the direct and reflected pulse. 

In this work, we propose an extension of the calibration setup by adding a second transmitter antenna to the setup to also determine the depth-dependence of the index-of-refraction profile. We determined the optimal positions of two emitters and showed that this setup outperforms current density-based measurements of the index-of-refraction profile. More importantly, the calibration system directly measures what is relevant for neutrino detection: the propagation times of radio waves for different trajectories through the ice, and does not rely on an empirical conversion of density to index-of-refraction. The setup even allows for continuous monitoring of the firn properties, e.g., a measurement every 12 hours. Equipping a few radio detector stations of a future array with this calibration system will be beneficial to confirm uniform and stable firn properties.

Based on the capabilities of the in-situ calibration system to reconstruct the snow accumulation to \SI{3}{mm} precision and the index-of-refraction profile more than 10 times more precise than current density-dependent reconstruction methods, we studied the impact of these uncertainties on the reconstruction on the neutrino vertex distance, neutrino direction, and neutrino energy. We found that the calibration system proposed here measures the ice properties well enough to have negligible impact on the reconstruction of the neutrino properties. We also studied much larger uncertainties in snow height and showed that even an uncertainty of \SI{10}{cm} is still tolerable. This alleviates the potential concern that local variations in snow height around the detector station -- if they exist -- will have a significant effect on the reconstruction of neutrino properties. 

\appendix

\section{Calibration System at Summit Station, Greenland}
\label{sec:greenland}

\begin{figure}[tbp]
        \centering
        \includegraphics[width = \textwidth]{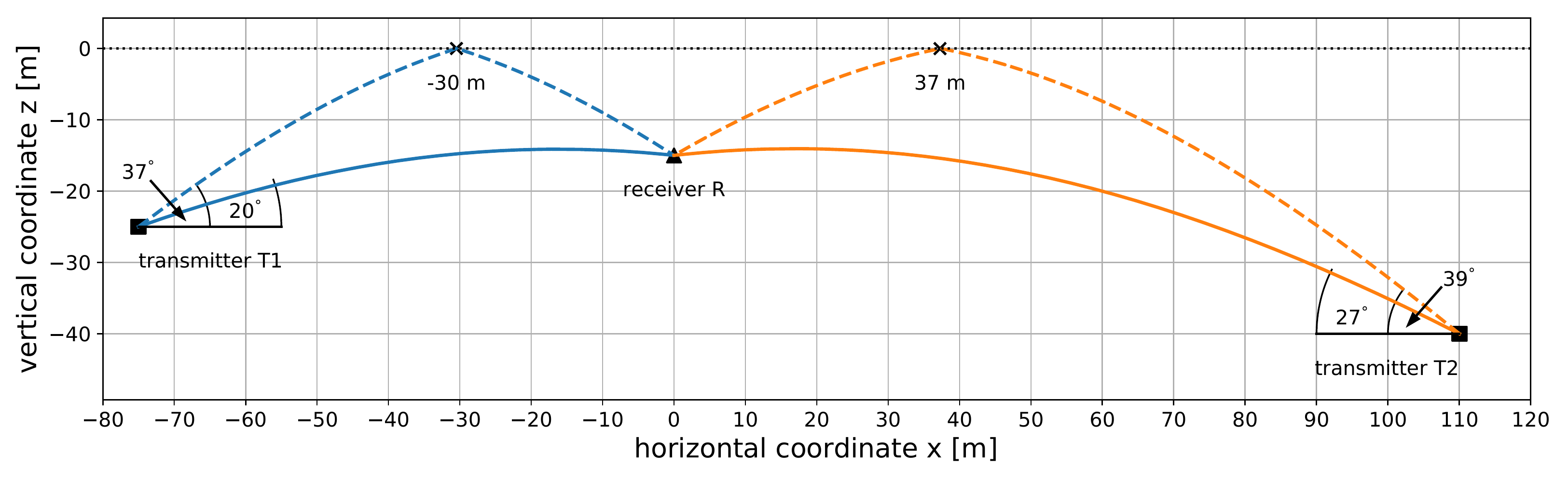}
        \caption{Sketch of the calibration system at Summit Station on Greenland consisting of two transmitters (T1 and T2, black square) and one receiver (R, black triangle) antenna. The optimised antenna configuration from Tab.~\ref{tab:greenland} is T1~=~[-75,~-25]~m and T2~=~[110,~-40]~m.}
        \label{fig:greenland}
    \end{figure}

Here, we report the optimal transmitter positions for the calibration system at the RNO-G site in Greenland, and the ARIANNA site on the Ross Ice Shelf, Antarctica. 
Our simulation study for the Greenland site finds $100 \times 99$ allowed transmitter configurations for depths between $-5$~m to $-40$~m. The optimal antenna  configuration is T1~=~[-75,~-25]~m and T2~=~[110,~-40]~m (Fig.~\ref{fig:greenland}) which minimizes the correlation while not significantly worsening the metric (Tab.~\ref{tab:greenland}). We determine the (relative) per-100-events statistical uncertainty to be $0.05\%$ in $\alpha$ and $0.04\%$ in $z_0$. For deeper stations up to \SI{-200}{m} we find T1~=~[-110,~-40]~m and T2~=~[310,~-200]~m, with a resolution of $0.03\%$ in $\alpha$, $0.02\%$ in $z_0$ and a correlation of $0.63\%$. Similarly, for stations of up to \SI{-20}{m} we obtain T1~=~[-55,~-15]~m and T2~=~[80,~-20]~m, with a resolution of $0.07\%$ in both $\alpha$ and $z_0$ with a correlation of $4.6\%$. We repeat the calculations for the systematic uncertainty from antenna deployment for the medium depth setup ($\leq$ \SI{-40}{m}) and find $0.04\%$ in $\alpha$ and $0.09\%$ in $z_0$ for an antenna displacement of \SI{1}{cm}, which is very similar to the results obtained for the South Pole site. We find that for Greenland, the statistical uncertainty in $\alpha$ is larger than the systematic uncertainty, whereas the uncertainty in $z_0$ is dominated by the systematic uncertainty.

\begin{table}[tbp]
        \centering
        \smallskip
        \begin{tabular}{cc|cccc}
            \hline\hline
            \multicolumn{6}{c}{metric-optimised configuration}\\
            \hline
            T1 & T2 & $\mathcal{M}$ & $\rho_{\alpha, z_0}$ & $\sigma_{\alpha}/\alpha^\mathrm{true}$ & $\sigma_{z_0}/z_0^\mathrm{true}$ \\
            $[-90,\ -25]$~m & $[110,\ -40]$~m & $7.1\cdot 10^{-5}$ & 49$\%$ & 0.08$\%$ & 0.05$\%$\\ 
            \hline\hline
            \multicolumn{6}{c}{|correlation| < 5$\%$ + metric-optimised configuration}\\
            \hline
            T1 & T2 & $\mathcal{M}$ & $\rho_{\alpha, z_0}$& $\sigma_{\alpha}/\alpha^\mathrm{true}$ & $\sigma_{z_0}/z_0^\mathrm{true}$ \\
            $[-75,\ -25]$~m & $[110,\ -40]$~m & $7.5\cdot 10^{-5}$ & 1.8$\%$ & 0.05$\%$ & 0.04$\%$\\ 
            \hline\hline
        \end{tabular}
        \caption{\label{tab:greenland} Optimal pair and corresponding metric, correlation and per-100-events parameter uncertainty at Greenland for the metric-optimised pair in the upper part and correlation-constrained, metric-optimised pair in the lower part.}
    \end{table}

\section{Calibration System at Moore's Bay on the Ross Ice Shelf}
\label{sec:mooresbay}
\begin{figure}[tbp]
        \centering
        \includegraphics[width = \textwidth]{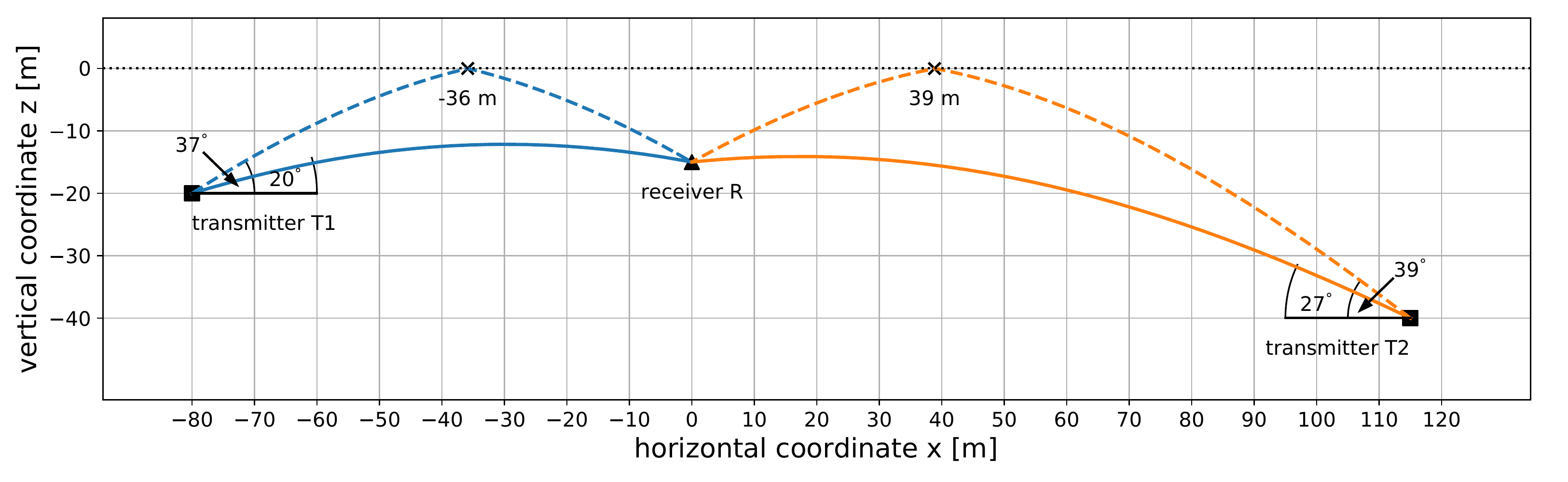}
        \caption{Sketch of the calibration system at the Ross Ice Shelf on Moore's Bay consisting of two transmitters (T1 and T2) and one receiver (R) antenna. The optimised antenna configuration from Tab.~\ref{tab:mooresbay} is T1~=~[-80,~-20]~m and T2~=~[115,~-40]~m.}
        \label{fig:mooresbay}
    \end{figure}

For the Moore's Bay site, we find $116 \times 115$ allowed transmitter configurations for depths between $-5$~m to $-40$~m. The optimal antenna configuration is T1~=~[-80,~-20]~m and T2~=~[115,~-40]~m (Fig.~\ref{fig:mooresbay}) that minimizes the correlation while not significantly worsening the metric (Tab.~\ref{tab:mooresbay}). We determine the (relative) per-100-events statistical uncertainty to be $0.06\%$ in $\alpha$ and $0.03\%$ in $z_0$. For deeper stations up to \SI{-200}{m} we find T1~=~[-105,~-30]~m and T2~=~[330,~-200]~m, with a resolution of $0.03\%$ in $\alpha$, $0.02\%$ in $z_0$ and a correlation of $1.8\%$. Similarly, for stations of up to \SI{-20}{m} we obtain T1~=~[-55,~-15]~m and T2~=~[80,~-20]~m, with a resolution of $0.07\%$ in both $\alpha$ and $z_0$ and a correlation of $4.1\%$.  We repeat the calculations for the systematic uncertainty from antenna deployment for the medium depth setup ($\leq$ \SI{-40}{m}) and find $0.04\%$ in $\alpha$ and $0.09\%$ in $z_0$ for an antenna displacement of \SI{1}{cm}, which is very similar to the results obtained for the South Pole site. We find that for Moore's Bay, the statistical uncertainty in $\alpha$ is larger than the systematic uncertainty, whereas the uncertainty in $z_0$ is dominated by the systematic uncertainty.

\begin{table}[tbp]
        \centering
        \smallskip
        \begin{tabular}{cc|cccc}
            \hline\hline
            \multicolumn{6}{c}{metric-optimised configuration}\\
            \hline
            T1 & T2 & $\mathcal{M}$ & $\rho_{\alpha, z_0}$ & $\sigma_{\alpha}/\alpha^\mathrm{true}$ & $\sigma_{z_0}/z_0^\mathrm{true}$ \\
            $[-85,\ -20]$~m & $[115,\ -40]$~m & $7.6\cdot 10^{-5}$ & 20$\%$ & 0.06$\%$ & 0.04$\%$\\ 
            \hline\hline
            \multicolumn{6}{c}{|correlation| < 5$\%$ + metric-optimised configuration}\\
            \hline
            T1 & T2 & $\mathcal{M}$ & $\rho_{\alpha, z_0}$ & $\sigma_{\alpha}/\alpha^\mathrm{true}$ & $\sigma_{z_0}/z_0^\mathrm{true}$ \\
            $[-80,\ -20]$~m & $[115,\ -40]$~m & $7.7\cdot 10^{-5}$ & 3.9$\%$ & 0.06$\%$ & 0.03$\%$\\ 
            \hline\hline
        \end{tabular}
        \caption{\label{tab:mooresbay} Optimal pair and corresponding metric, correlation and per-100-events parameter uncertainty at Moore's Bay for the metric-optimised pair in the upper part and correlation-constrained, metric-optimised pair in the lower part.}
    \end{table}
    
\section*{Acknowledgements}
We thank the ARIANNA collaboration and especially Steven Barwick for feedback on the analysis and manuscript. We thank the radio working group of IceCube-Gen2 for their feedback on the analysis. The computations and data handling were enabled by resources provided by the Swedish National Infrastructure for Computing (SNIC) at UPPMAX partially funded by the Swedish Research Council through grant agreement no. 2018-05973.

\bibliographystyle{JHEP}
\bibliography{bib}

\end{document}